# Investigation of the Nature of the B[e] Star CI Cam in the Optical Range

E. A. Barsukova[1]*, A. N. Burenkov[1], V. P. Goranskij[2,1], S. V. Zharikov[3,4], L. Iliev[5], N. Manset[6], N. V. Metlova[2], A. S. Miroshnichenko[7,8,9,4], A. V. Moiseeva[1], P. L. Nedialkov[10], E. A. Semenko[1,11], K. Stoyanov[5], and I. A. Yakunin[1,12]

[1]*Special Astrophysical Observatory, Russian Academy of Sciences, Nizhnii Arkhyz, 369167 Russia*
[2]*Sternberg Astronomical Institute, Moscow State University, Moscow, 119234 Russia*
[3]*Institute of Astronomy, National Autonomous University of Mexico, Ensenada, Baja California, 22800 Mexico*
[4]*Al-Farabi Kazakh National University, Almaty, 050040 Kazakhstan*
[5]*Institute of Astronomy and National Astronomical Observatory Rozhen, Bulgarian Academy of Sciences, Sofia, 1784 Bulgaria*
[6]*Canadian-French-Hawaiian Telescope Corporation (CFHT), Kamuela, Hawaii, 96743 USA*
[7]*University of North Carolina at Greensboro, Greensboro, 27402 USA*
[8]*Central (Pulkovo) Astronomical Observatory, Russian Academy of Sciences, St. Petersburg, 196140 Russia*
[9]*Fesenkov Astrophysical Institute, Almaty, 050020 Kazakhstan*
[10]*Department of Astronomy, Sofia University, Sofia, 1504 Bulgaria*
[11]*National Astronomical Research Institute of Thailand, Mae Rim, Chiang Mai, 50180 Thailand*
[12]*St. Petersburg University, St. Petersburg, 199034 Russia*

**Abstract** − We report the results of 24 years of photometric and spectroscopic monitoring of CI Cam since its outburst in 1998. The aim of this study was to investigate CI Cam (MWC 84, XTE J0421+560), which flared up in April 1998 across all ranges of the electromagnetic spectrum. This investigation is crucial for understanding the B[e] phenomenon and the nature of this unique explosion. We utilized optical photometry to monitor the system's brightness in several filters, as well as medium- and high-resolution spectroscopy. In the early years of our research, we identified a system component responsible for the emission of the He II 4686 Å line, which moves in an elliptical orbit with a period of 19.407 days and an eccentricity of 0.44 - 0.49. Variations in optical brightness with the same period were observed, with an average amplitude of $0^m.04$. The total amplitude of the He II radial velocity variations was approximately 380 km s$^{-1}$. The equivalent width of the line varied on timescales of tens of minutes as well as with the orbital period, reaching maximum values when the companion passed the descending node of the orbit. The intensity of the He II 4686 Å emission has gradually increased over time. Slow radial velocity variations on a scale of decades were detected via high-resolution spectroscopy in the iron emission lines and a forbidden nitrogen line formed in the circumstellar nebula. We discussed three hypotheses for the acceleration in the stellar wind. Unfortunately, the available high-resolution spectra were insufficient to reach a definitive conclusion. Our photometric monitoring revealed pulsations of the main B component of the CI Cam system. Between 2005 and 2009, the B star exhibited multiperiodic pulsations with dominant periods of 0.5223, 0.41539, and 0.26630 days. However, since 2012, it pulsated in a single mode with a variable period in the $0^d.403-0^d.408$ range, depending on the star's luminosity. We interpret the pulsations from 2005 - 2009 as a resonance of radial modes, with the residual stable mode being the first overtone. The pulsations are coherent over several months, with average amplitudes of $0^m.02-0^m.04$ in the $V$ band. We applied Eddington's formula for the pulsation constant and Christy's computations for RR Lyrae-type stars to refine the classification of the pulsating component of CI Cam and its parameters. The pulsation data constrain the spectral type of the primary component to B0-B2 III, the distance to the system to 2.5-4.5 kpc, and the absolute visual magnitude ($M_V$) to the range of $-3^m.7$ to $-4^m.9$. The classification of the main component of CI Cam as a sgB[e] is ruled out due to the observed pulsation periods. CI Cam is likely a system in the post-first mass exchange stage and may belong to the FS CMa-type group of objects exhibiting the B[e] phenomenon.

## 1. Introduction

B-type stars with forbidden emission lines in their spectra were first identified and described as peculiar Be-type stars with an infrared excess by Allen and Swings (1976). They included CI Cam in their list of such objects under the name MWC 84. At the IAU Symposium 70 in 1976, P. S. Conti proposed the designation B[e] for these stars. Based on a twenty-year investigation by Lamers et al. (1998), the following spectral features were established as classification criteria for B[e]-type stars:
- strong Balmer emission lines;
- low excitation permitted emission lines of predominantly low ionization metals in the optical spectrum, e.g., Fe II;
- forbidden emission lines [Fe II] and [O I] in the optical spectrum;
- a strong emission excess in the near and mid-infrared (IR) regions from hot circumstellar dust.

These features are observed in stars of various masses and luminosities at different evolutionary stages. Lamers et al. (1998) categorized these stars into five groups:
1) HAeB[e]—Herbig pre-main sequence B[e]-type stars,
2) symbB[e]—symbiotic stars with a cool giant and a white dwarf (or a neutron star),
3) cPNB[e]—compact planetary nebulae,
4) sgB[e]—supergiants,
5) unclB[e]—unclassified objects.

The nature and evolutionary status of the unclassified group remained unclear. Later, Miroshnichenko (2007) isolated a new group of dust-forming, non-supergiant B[e] stars among the unclassified objects— termed FS CMa-type stars. These are thought to be binary systems in the fast mass exchange phase (or having recently passed it), possibly accompanied by partial mass loss and the formation of circumstellar dust.

CI Cam (MWC 84) was discovered by Merrill and Burwell (1933) as a star with strong H, He, and Fe II emissions. The first spectra of the star were published by Downes (1984) and Miroshnichenko (1995), with *UBVRIJHK* photometry reported by Bergner et al. (1995). Based on these data, the star was classified as B0V. Brightness variations with an amplitude of $0^m.4$ were discovered, along with an infrared excess initially attributed to a cool K0 II star. Consequently, the star was classified as SymbB[e] and listed in the General Catalogue of Variable Stars (GCVS) under the Z And-type classification (ZAND).

Despite extensive studies of CI Cam across various wavelengths and using different methods, several crucial questions remain unanswered. Uncertainties remain regarding the distance estimate, the properties of the disk surrounding the system, the specific B[e]-star group membership, and the evolutionary stage of the system. Before the outburst on April 1, 1998, the classification of the star was debatable. Following the outburst, CI Cam was considered a candidate for sgB[e] supergiants and even ULX-type objects (Ultraluminous X-ray sources) and supernova impostors. The powerful outburst across a wide wavelength range, along with the pulsations of the main system component that we discovered, makes CI Cam unique among stars of this type. Studying this object may enhance our understanding of the evolution of multiple systems and the B[e] phenomenon.

## 2. Outburst and Unique Properties of the CI Cam System

In early April 1998, a powerful X-ray outburst was detected by RXTE/ASM, leading to the designation of the source as XTE J0421+560, which was identified with CI Cam (Smith et al., 1998). The flux level continued to rise for several hours, reaching its maximum on April 1 at $0^h 57^m$ UT (HJD 2450904.54). The X-ray spectrum was relatively soft compared to typical X-ray novae and did not extend into the high-energy region above 60 keV (Belloni et al. 1999). A radio source appeared on April 1, with its flux peaking on April 3 at $20^h$ UT, showing a delay of $2^d.8$ relative to the X-ray peak. Mioduszewski and Rupen (2004) observed the radio remnant of the outburst using the VLBA, revealing an expanding bipolar structure similar to a shock wave passing through a dense circumstellar medium. The highest optical brightness was recorded with a 1.06-day delay relative to the X-ray peak, reaching $7^m.1$ in the *R* band (Hjellming et al., 1998). In a quiescent state, the *R* band brightness of the object is usually around $10^m.6$. During the outburst, the optical spectrum exhibited a strengthening of the emission lines visible during quiescence (Barsukova et al., 1998; Clark et al., 2000). The He II 4686 Å emission line showed the largest increase in equivalent width, reaching up to *EW* = 176 Å.

After the outburst, this parameter did not exceed 0.5 Å, or the line was not visible at all. However, the flux of the forbidden [N II] 5755 Å line remained constant throughout the outburst (Barsukova et al., 2002). The X-ray outburst lasted for 10–15 days. In the optical range, the increased brightness was observed until the end of the visibility season (JD 2450942, $V = 11^m.52$), persisting for more than 38 days. By the start of the next season, the brightness had returned to its quiescent level (JD 2451051, $V = 11^m.69$). Additionally, radio emission was detected throughout the entire next season, up to JD 2451300.

The 1998 outburst was interpreted as a thermonuclear explosion on the surface of a white dwarf in the CI Cam system (Ishida et al., 2004; Orlandini et al., 2000). The hard spectrum obtained by the orbital X-ray telescope (XRT) ASCA on April 3–4, 1998, is represented by a two-temperature optically thin thermal emission model (1.1 and 5.7 keV). A neutral iron emission was detected in the spectrum at 6.41 keV with an equivalent width of $EW = 90$ eV. A thermonuclear explosion on a white dwarf's surface is the mechanism for classical nova explosions. However, Clark et al., (1999), Hynes et al., (2002), and Robinson et al., (2002) considered supercritical accretion events onto a black hole or a neutron star in a high-mass X-ray binary (HMXB) as a possible cause of the explosion. The hard X-ray spectrum, which can be modeled by a power-law energy distribution, is also typical for X-ray transients with neutron stars and black holes. However, the fast variations in emission flux that are typical for such transients during outbursts were not observed in CI Cam (Ishida et al., 2004).

The distance to CI Cam remains undetermined and is estimated differently in various studies, ranging from 1 to 17 kpc. However, estimates of the B-type star's luminosity class, as well as the interpretation and classification of the CI Cam system and the 1998 event, are contingent upon the distance estimate. Conflicting data also emerge from the Gaia orbital observatory data releases. The distances derived by direct conversion from parallaxes are as follows: 1.41 kpc +0.72/-0.36 kpc (DR1); 10.92 kpc +6.65/-2.98 kpc (DR2); 4.76 kpc +0.36/-0.32 kpc (DR3). Thus, the distance estimates fall within the range from 1 to 17 kpc. According to the latest Gaia EDR3 data, with statistical corrections taken into account, the distance to CI Cam is estimated as 4.1 kpc +0.3/-0.2 kpc.

The optical component of CI Cam, based on data from Clark et al., (1999), Hynes et al., (2002), and Robinson et al., (2002), was initially attributed to a B0–B3-type supergiant. High-order Balmer lines' wide absorption wings were proposed by Hynes et al., (2002) for its spectral classification. These wings were most noticeable in the Hδ and Hε line profiles and could potentially belong to the photospheric spectrum of a B star. However, this diagnostic was not utilized due to significant noise contamination of these line profiles and the star's continuum by numerous wind emission lines. Based on spectral and photometric data, the authors classified the optical component as sgB[e] with a probable B0–B2 spectral type and a luminosity of at least $10^{5.4} L_\odot$, placing CI Cam among the hottest stars with the highest luminosities. Using spectral data for normal stars from collections of spectra by Jacoby et al., (1984) and Le Borgne et al., (2003), Barsukova et al., (2006a) analyzed the full width at zero intensity (*FWZI*) of the Hδ and Hη lines at the continuum level using spectral data for normal B-stars of various luminosity classes and CI Cam. It was found that the sequences for B-type stars with Iab luminosity class and III–V class stars do not overlap on the "spectral type–full width" (Sp–*FWZI*) diagram, and CI Cam is not located among supergiants (Barsukova et al., 2006a). These data were used to classify the object as B4 III–V. However, Miroshnichenko et al., (2002) did not detect any B-star photospheric lines in CI Cam, even in high-resolution spectra.

Further regular spectral and photometric observations of CI Cam in the optical range during quiescence after the outburst revealed many unique properties of this system. The spectrum exhibits shifts in the He II 4686 Å line with a period of 19.407 days, indicating an emission source moving on an elliptical orbit (Barsukova et al., 2006a,b). The orbital period was also detected in the photometric data. High-resolution spectra revealed a slow drift with an amplitude of up to 14 km s$^{-1}$ in the B-star wind line radial velocities, which could imply an accelerated motion of an extended nebula as a whole (Barsukova et al., 2007). Subsequent observations from 2007 to 2016 showed that the direction of this drift unexpectedly changed to the opposite in a jump-like manner. This phenomenon was explained by a close transit of a third component moving on an elliptical orbit. A close transit of such a component in 2007–2008 could have caused the subsequently observed increase in the star's brightness (Goranskij et al., 2017).

Šimon et al., (2007a,b) discovered a fast brightness variation with an amplitude of 0.02 magnitudes, similar to the light variations of the microquasar LS 5039. Observations by Barsukova and Goranskij (Barsukova and Goranskij 2008, 2009; Goranskij and Barsukova 2009) indicated that these are double-mode pulsations of a B4 III–V optical component. While pulsations occur in B and Be stars, this is a unique case among B[e] stars. Observations in 2018 revealed the star's transition to single-mode pulsations with a period of 0.4062 ± 0.0015 days, which was accompanied by an increase in the star's *V*-band brightness by $0^m.3$. Spectral monitoring over several nights showed rapid variability in the equivalent width of the He II 4686 Å line on a scale of tens of minutes, ranging from zero to 1.15 Å. Moreover, an emission outburst up to *EW* = 1.9 Å (Barsukova et al., 2021) was observed during the passage of the He II emission source through the descending node of the orbit.

Thureau et al. (2009) used interferometry in the IR range to detect a dust ring around the B star with an external radius of *a* = 7.58±0.24 milliarcseconds, with its axis tilted towards the line of sight by an angle of *i* = 67°. Observations were carried out with the Infrared Optical Telescope Array (IOTA) in Arizona, USA, and the Palomar Testbed Interferometer (PTI) in California, USA. If we adopt a distance to CI Cam of 5 kpc, the size of the major axis of this structure is 37.9±1.2 AU (8150 $R_\odot$). Assuming the orbit determined by Barsukova et al. (2006a) is coplanar with the dust ring, its semi-major axis would be 0.35 AU ($75 R_\odot$). Such an orbit is located deep within the ring and is unresolvable by these interferometers. No traces of a massive third component, as reported by Barsukova et al. (2007), were detected, although the search for this component is complicated due to asymmetric dust emission.

The results of a study of the X-ray spectrum and variability of CI Cam in a quiescent state over a period of 40,000 seconds with the XRT on XMM-Newton on February 24, 2003 (HJD 2452695.0243–2452695.7436) are presented in Bartlett et al., (2013). The spectrum was obtained near the exterior conjunction phase of the He II emission source. The spectrum in the 3–12 keV range is a power-law with strong absorption, typical in terms of photon index for X-ray systems with a Be star and a neutron star. The authors concluded that the spectrum cannot be represented by the white dwarf model from Ishida et al., (2004). Bartlett et al., (2019) studied the X-ray spectra of CI Cam and the emission flux variations of the object in this range in a quiescent state on five occasions using XMM/EPIC-pn, NuSTAR/FMPA and FMPB, and multiple X-ray images from Swift XRT obtained over 150 days. Additionally, the 1.2-meter Mercator telescope (La Palma, Spain) was used to obtain optical spectra on October 19 and 23, 2016, with the HERMES echelle spectrograph at a resolution of *R*~85000. The X-ray spectrum was interpreted as a sum of a power-law spectrum from the compact object with accretion and a reprocessed component of this spectrum that dominates the continuum at energies less than 3 keV and is the source of the iron line. The latter may be represented by both a black body and a power-law distribution. It is formed in a dense circumstellar medium, which is a significant absorption source in the X-ray region. In a quiescent state, CI Cam is highly variable in X-rays on a scale of one day. Both the integral flux and the spectral energy distribution vary, although every spectrum exhibits the same components. The changes in the spectrum were explained by the authors as due to both the cloudy structure of the surroundings and the change in accretion rate. Swift data showed variations on a scale of approximately 75–100 days. However, no evidence was found for either the 19.4-day period or any pulsations. According to Bartlett et al. (2019), the He II 4686 Å emission is completely absent in the Mercator/HERMES spectra. The authors conclude that due to the unusual variability of this emission, caution should be exercised when considering the $19^d.4$ period reported by Barsukova et al. (2006a). Bartlett et al. (2019) cite other studies that suggest "the transient nature of this line strongly suggests that it is produced in the stellar wind, rather than in the star itself, making it unsuitable for tracing dynamics."

However, the most remarkable conclusion of Bartlett et al. (2019) is the similarity of CI Cam in its X-ray and optical spectrum, as well as in its behavior during outburst and quiescent states, with the well-known object ULX-1, sgB[e], and the supernova impostor SN2010da in the NGC 300 galaxy. We confirmed this similarity in optical spectra by comparing the spectra of SN2010da taken 2 and 15 days after the outburst peak (Villar et al. 2016) and those of CI Cam taken 3–5 days after the outburst peak by Barsukova et al. (1998). The optical component of SN2010da is defined in Bartlett et al. (2019) as an sgB[e] star with a "moderate luminosity of the order of $10^4 \, L_\odot$". Heida et al. (2019) discovered that in the near IR range of the SN2010da spectrum, absorption from a cool star dominates wavelengths greater than 700 nm. The absorption features, particularly the CO molecular band heads of a red supergiant (RSG), are conspicuous in the 1575–1700 nm region. The identity of the RSG and ULX-1 is established very reliably, with the RSG parameters being $R = 310 \pm 70 \, R_\odot$ and $M = 8–10 \, M_\odot$. So, the spectrum of a B[e] supergiant may also be formed without a B supergiant. Multiple emission lines in the spectrum that were attributed to sgB[e] have nothing to do with the accretion donor. Earlier,

Carpano et al. (2018) discovered pulsations of SN2010da in the X-ray region with a period of 31.6 seconds, which are attributed to an accreting neutron star. The presence of a strong stellar wind in SN2010da is probably due to an accretion disk around the compact component in this binary system. Model grids for such winds are presented in Kostenkov et al., (2020). However, the assumption that the 1998 CI Cam outburst identifies it with SN impostors is of special interest and demands verification by observational data.

In this work, we present a survey and analysis of optical observations of CI Cam obtained over the course of 24 years in a quiescent state after the 1998 outburst. The results of photometric and spectral studies in the optical region are found to be more informative than those in other domains of the electromagnetic spectrum.

## 3. Photometry

The photometric study of CI Cam in a quiescent state presented in this work is based on observations carried out between August 26, 1998 (JD 2451051, 147 days after the X-ray outburst maximum), and May 21, 2022 (JD 2459721). These observations include multicolor data in the Johnson *UBVRI* system and Cousins *RI*, as well as campaigns of long-term continuous monitoring in the *V* band over periods ranging from 1 to 3 weeks, with up to 12 hours of observation per night. A detailed list of observatories, telescopes, and instruments is provided in Table 1. The primary series of observations were conducted by N. V. Metlova using V. M. Lyuty's *UBV*- electrophotometer with the 60-cm Zeiss telescope at the Crimean SAI MSU station. These observations form a homogeneous series spanning 21 years, with other *UBV* observations adjusted using corrections determined from simultaneous or nearby observations. Another consistent *UBV* $(RI)_C$ series covering a duration of 19 years (2003–2022) was obtained with the 1-meter Zeiss telescope at SAO RAS, equipped with a CCD photometer. All other CCD observations in $R_C$ and $R_J$ filters were standardized to this system for consistency. Although observations in $I_C$ and $I_J$ bands were conducted to a lesser extent, we were unable to standardize them to a single system and thus did not consider them in this paper. The average accuracy of the *BVR* photoelectric and CCD observations ranges from 0.01 to 0.03 magnitudes, while in the *U* band, it is up to 0.05 magnitudes. *V*-band observations conducted with the AZT-5 telescope at the Crimean SAI MSU station in monitoring mode have accuracies ranging from 0.002 to 0.004 magnitudes. The AZT-5 telescope is a Maksutov meniscus telescope with the CCD mounted at the prime focus. A dedicated investigation has demonstrated that the higher measurement accuracy at the prime focus of the AZT-5 due to the absence of blending and photometer parts with surfaces parallel to the telescope's optical axis, consequently eliminating slanting reflections when acquiring field calibration frames.

GSC 3723.0054 ($V = 10^m.401$, $B-V = 0^m.759$, $U-B = 0^m.336$) was used as a comparison star for photoelectric observations. The check star was GSC 3723.0104, with $V = 12^m.386$, $B-V = 0^m.617$, and $U-B = 0^m.408$. These same stars were used in Barsukova et al. (2002) and Barsukova et al. (2006a). For CCD observations, GSC 3723.0080 was used as a comparison star, with $V = 12^m.387$, $B-V = 0^m.771$, $U-B = 0^m.218$, $V-R_C = 0^m.457$, and $(R-I)_C = 0^m.442$. The magnitudes were obtained by Henden and Munari (2006). All three check stars chosen for CCD observations turned out to be variable (Goranskij and Barsukova 2007) and are designated in the GCVS as V338 Cam (LB type, corresponding to slow irregular late spectral type variables, Ampl. = $0^m.82$ *V*); V339 Cam (GSC 3723.0104, δ Sct type, Ampl. = $0^m.033$ *V*); and V340 Cam (GSC 3723.0602, eclipsing W Uma-type, Ampl. = $0^m.065$ *V*). Flat field, dark, and bias (zero level shift) frames were used for standard calibration, with the exception of *UBVRI* photometer data obtained with the Zeiss-1000 telescope of SAO RAS, where the CCD chip is cooled to a temperature of −130°C (and dark frames were not used). Stellar magnitudes were determined using the WinFITS[1] software in aperture mode. The program is capable of manual and automated cleaning of stellar profile images from defects, close stellar components, and cosmic ray hits. Observations of CI Cam continue, and as of the date of preparing this paper, our collection includes 16,044 observations.

For the analysis, we also used the NASA TESS (Transiting Exoplanet Survey Satellite) photometry results for CI Cam, which are openly accessible in the MAST archive of the Space Telescope Science Institute (STScI), USA[2]. The TESS detector response curve spans a wavelength interval from 6000 Å to 10,000 Å, and its center coincides with the center of the $I_C$ band at λ 7865 Å.

---

[1] Developed by V. P. Goranskij. [2] https://outerspace.stsci.edu/display/TESS/TESS+Holdings+Available+by+MAST+Service

The accuracy of the CI Cam brightness measurement with the TESS satellite is between 0.0005 and 0.0010 magnitudes. The TESS data contain 17308 measurements over a 25-day interval (JD 2458816.09–2458841.15), with a single one-day interruption.

Table 1. Photometric observations of CI Cam

| JD 2450000+ start–end | Number of observations | System | Instrument/CCD | Archive table mark |
|---|---|---|---|---|
| 100-cm Zeiss (SAO RAS) | | | | |
| 1928–9721[a,b] | 319 | $UBVR_CI_C$ | CCD EEV 42-40 | SO |
| 8459–9679[a] | 23 | $UBVR_CI_C$ | MMPP CCD E2V 42-40 | MM |
| 70-cm AZT-2 (SAI MSU) | | | | |
| 1143–1235[a] | 28 | $BVR_J$ | phe SgV PTM | f7 |
| 60-cm Zeiss (CAS SAI MSU) | | | | |
| 1103–8610[c] | 759 | $UBV$ | phe Lyuty PTM | Me |
| 1192–3740[a] | 118 | $BVR_J$ | CCD SBIG ST7 | C7 |
| 2022–3036[a] | 3 | $BVR_J$ | CCD Apogee 7 | A7 |
| 3355–4085[a] | 24 | $BVR_C$ | CCD PI VersArray | VA |
| 6292–6350[a] | 5 | $UBVR_J$ | CCD Apogee 47 | 47 |
| 60-cm Zeiss (Stara Lesna, SK) | | | | |
| 4778–4788[d] | 50 | $V$ | CCD ST10MXE | SH |
| 50-cm AZT-5 (CAS SAI MSU) | | | | |
| 3582–4814[a,b] | 4163 | $V$ | CCD Meade Pictor 416 | Pi |
| 6267–6288[a,b] | 1176 | $V$ | CCD Apogee Alta U8300 | AA |
| 6889–9413[a,b] | 8815 | $UBVR_CI_C$ | CCD Apogee Alta U8300 | A5 |
| 30-cm Zeiss guide (SAO RAS) | | | | |
| 9510–9721[a] | 488 | $r$ | CCD Atik-414x | Ak |

Observers: [a]—V.P. Goranskij, [b]—E.A. Barsukova, [c]—N.V. Metlova, [d]—S.Yu. Shugarov.

The CI Cam light curves in a quiescent state in the $UBVR_C$ filters are shown in Fig. 1. Monitoring dates are marked beneath the $V$-band light curve by letters $a$–$h$. The results of monitoring — the phase light curves with pulsation periods determined by frequency analysis for each series — are marked by the same letters in Fig. 2. These figures also show the TESS observation dates and the phase curve.

The light curves show variations on time scales from tens of minutes to several years, including short-term outbursts and brightness decays. However, the most dominant feature is the outburst with a peak in 2013, preceded by a gradual brightness increase starting in 2009. A slow decline is observed after the maximum over four years, after which an increased brightness level settled in, approximately 0.25 magnitudes higher than the average brightness level before the outburst in all filters. The steepest part of the brightness increase before the maximum lasted for 200 days. In this part of the curve we obtained photometric series with a duration of 21 days, suitable for studying pulsations. Fig. 3a–d shows the $V$-band light curve and the color index curves for $U-B$, $B-V$ and $V-R_C$. While indices $U-B$ and $B-V$ either change little during the described light variations or do not change at all, the $V-R_C$ index has notably decreased by 0.1 magnitudes. The position of CI Cam on the $(U-B)$–$(B-V)$ color-color diagram was determined by photoelectric $UBV$ observations (observer N. V. Metlova). Strong emissions of the surrounding gas envelope dominate in the CI Cam spectrum. To determine the position of the B star itself from its continuum, we determined and subtracted the contribution of the emissions in each filter. The results are depicted in Fig. 4. Notably, CI Cam is positioned above the black body line representing the highest temperature within the region hosting active galactic nuclei, quasars, and cataclysmic variables with highly luminous accretion disks. This observation might be linked to the presence of Balmer continuum emission in the star's ultraviolet spectrum. However, it's essential to note that this continuum was not subtracted when considering the emission spectrum's contribution. To analyze the periodic components of CI Cam's light curves, we employed the EFFECT software developed by V.P. Goranskij. Orbital period refinement utilized the Fourier transform method applied to a discrete time series, employing the Deeming technique (Deeming 1975) on day-averaged observations. Pulsation periods were determined and refined through intensive tracking series. Simultaneously, both the orbital period and slow irregular light variations typical of CI Cam were eliminated through Fourier decomposition into periodic components and the "prewhitening" procedure.

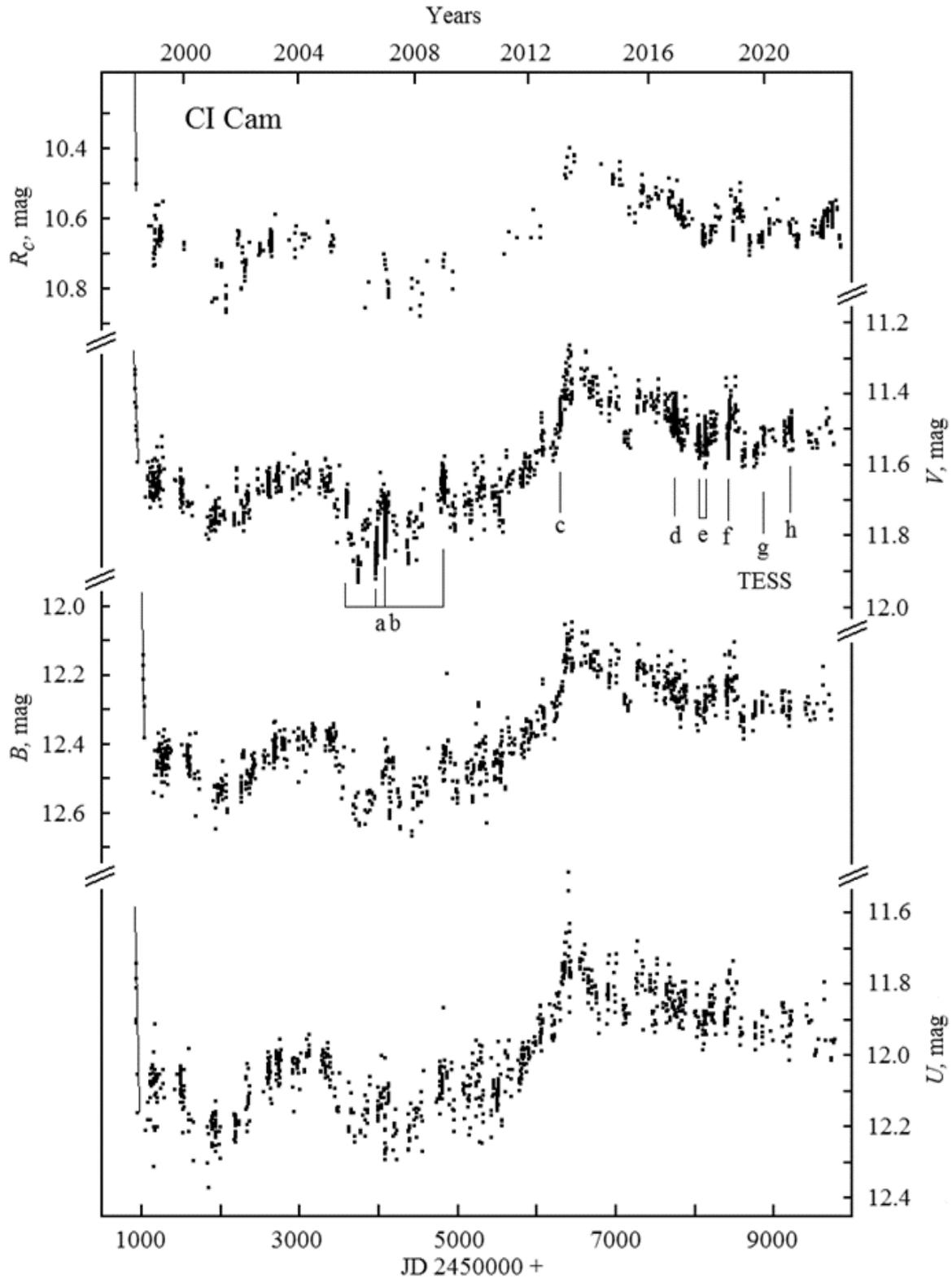

Fig. 1. *UBVR$_C$ light curves for CI Cam in a quiescent state following the 1998 outburst (from bottom to top). The solid lines on the left indicate observations during the brightness decline after the outburst. Intensive monitoring campaigns for rapid light variations and pulsations are marked on the V-band light curve with indicators and the letters a−h, where g denotes the TESS observations.*

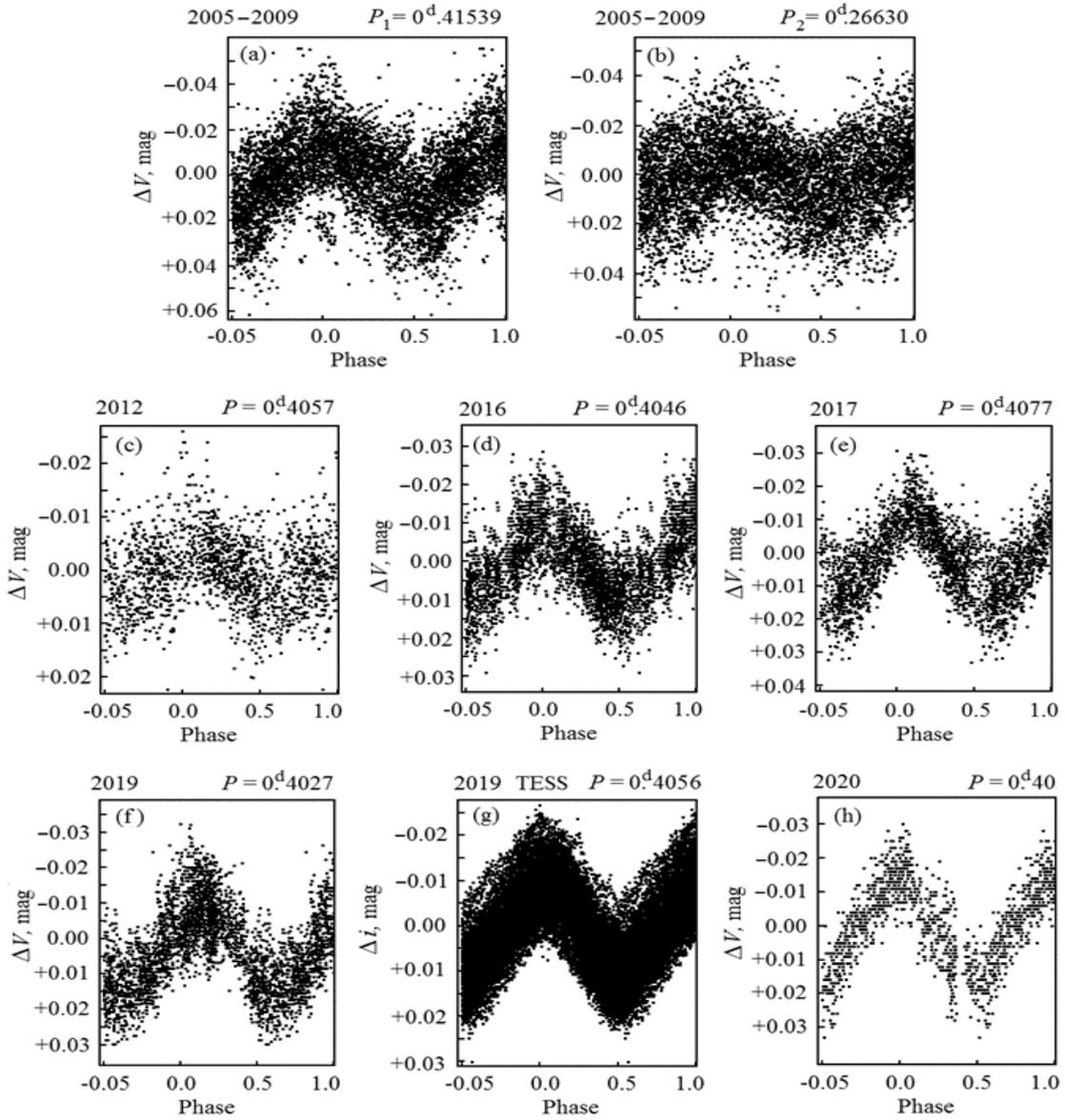

Fig. 2. *Pulsations of CI Cam revealed through monitoring. (a, b) Light curves in the first overtone $P_{1H}$ ($P_1$) and in the radial mode ($P_2$) with a period ratio of $Px/P_{1H} = 0.64$. (c)–(h) Pulsations in the first overtone ($P_{1H}$) during the period from 2012 to 2020; (g) shows the light curve from TESS observations.*

## 4. Spectroscopy

Most of the low- and medium-resolution spectra were obtained with the Zeiss-1000 telescope of SAO RAS, equipped with the UAGS spectrograph, as part of the study of the orbital motion of the He II 4686 Å line emission source. Diffraction gratings R325, R400, R651, and 1302 were used. The spectra taken with gratings R325 and R400 cover the entire optical range, while those with R651 cover the 3510–6250 Å wavelength range. Grating 1302 was primarily used to observe the 3860–5272 Å spectral region, with a resolution of 1.8–2.8 Å. This is the most suitable UAGS configuration for monitoring variations in the He II 4686 Å emission line, including profile shape, equivalent width (*EW*), and radial velocities. The spectral range was selected so that the He II line was positioned close to the maximum quantum efficiency of the device by wavelength. For wavelength calibration, a Ne-Ar lamp spectrum was used. To control flexure, comparison spectra were obtained before and after the star's spectrum exposure or at least once per hour during monitoring. The spectral resolution was estimated using the full width at half maximum (*FWHM*) of the lamp spectrum line widths at the frame center. The accuracy of determining radial velocities was estimated by the root-mean-square deviation of the lamp spectrum line measurements from a model representation of the dispersion curve as an n-degree polynomial (n = 1– 4).

Table 2. Low- and medium-resolution spectral observations of CI Cam

| JD 2450000+ start–end | Range, Å | R, Å | Number of spectra | Grating |
|---|---|---|---|---|
| 6-meter BTA telescope of SAO RAS, UAGS | | | | |
| 1204–3328[a] | 3800–6990 | 4.3–5.3 | 5 | 1302 |
| 6-meter BTA telescope of SAO RAS, SCORPIO | | | | |
| 1204–9609[b] | 3624–5396 | 5.6–6.0 | 13 | VPHG1200G |
|  | 5651–7373 |  |  | VPHG1200R |
|  | 3615–7892 | 18.2 |  | VPHG550 |
| Zeiss-1000 SAO RAS, UAGS | | | | |
| 1485–9659[c] | 3476–7997 | 5.8–9.0 | 80 | R325, R400 |
|  | 3506–6708 | 4.0–6.7 | 6 | R651 |
|  | 3860–5275 | 1.8–2.8 | 187 | 1302 B |
|  | 5730–7125 | 1.5–2.1 | 19 | 1302 R |

Observers: [a]—A.N. Burenkov, N.V. Borisov, G.G. Valyavin, D.N. Monin, S.N. Fabrika, E.A. Barsukova, [b]—S.N. Fabrika, D.N. Monin, A.S. Vinokurov, A.F. Valeev, [c]—A.N. Burenkov, V.V. Vlasyuk, N.V. Borisov, V.P. Goranskij.

Table 3. High resolution spectral observations of CI Cam

| JD2450000+ start–end | Range, Å | R, Å | Number of spectra | Instrument |
|---|---|---|---|---|
| 6-meter BTA, SAO RAS | | | | |
| 1150–1150 | 3936–5676 | 0.3 | 1 | PFES |
| 1206–1206 | 5000–7700[a] | 0.23 | 1 | LYNX |
| 2300–5933 | 4465–5930[b] | 0.14 | 12 | NES |
| 9551–9612 | 4225–4981[c] | 0.25 | 2 | MSS |
| CFHT 3.6-m Hawaii | | | | |
| 3363–8438 | 3792–10492 | 0.09 | 2 | ESPaDOnS |
| 2.7-m McDonald | | | | |
| 2541–3725 | 3740–7523 | 0.10 | 5 | CS2 |
| 2.7-m San Pedro Martir | | | | |
| 5142–7303 | 3630–7327 | 0.26 | 9 | REOSC |
| 2-m RCC telescope Rozhen | | | | |
| 8718–9531 | 4154–8849[d] | 0.19 | 5 | Echelle |

Observers: [a]—E.A. Barsukova, D.N. Monin, [b]—V.G. Klochkova, V.E. Panchuk, M.V. Yushkin, [c]—I.A. Yakunin, V.P. Goranskij, [d]—K.A. Stoyanov, L.H. Iliev.

Spectra were processed using the ESO MIDAS environment (E. A. Barsukova) and the SPEC software package (V. P. Goranskij). The wavelength scale was calibrated primarily using averaged Ne-Ar lamp spectra obtained both before and after the acquisition of the object spectrum. Additionally, computed heliocentric corrections were applied to the derived radial velocities of the object. Part of the spectral material was processed by calibrating against the FeII wind emission lines in the spectrum of the CI Cam star itself. These lines exhibit a "rectangular" profile with steep edges (Klochkova et al., 2022; Miroshnichenko et al., 2002) and are approximately 75–100 km s$^{-1}$ wide, which determines the required measurement accuracy. Previously, we observed a slow drift of these lines in the high-resolution spectra of CI Cam within a 14 km s$^{-1}$ range (from −55 to −41 km s$^{-1}$; Barsukova et al., 2007) and analyzed the dependence of radial velocity on time. The amplitude of this motion is approximately equal to or significantly smaller than the accuracy of the He II radial velocity determination from our low- and medium-resolution spectra. By knowing the heliocentric radial velocity of the Fe II lines at the time of measuring the relative velocity of the He II emission, we computed the heliocentric velocity of He II. Such measurements are not affected by the bending of the spectrograph structure.

A small number of spectra were obtained with the 6-meter BTA telescope using the SCORPIO focal reducer (Afanasiev and Moiseev, 2005), equipped with volume phase holographic gratings VPHG 550G, VPHG 1200G, and VPHG 1200R. The data on the telescopes and equipment used for low- and medium-resolution spectroscopy are outlined in Table 2. Overall, 304 low- and medium-resolution spectra were obtained in the blue spectral region, of which 38 (about 12.5%) display no He II 4686 Å emission at all, or it is so faint that its radial velocity cannot be determined. The emission is also visible and measurable in 10 high-resolution spectra. The dependence of the equivalent width of the He II 4686 Å line on time is shown in Fig. 3(e).

To study the acceleration in the wind emission lines, we used high-resolution spectra. The main parameters of the telescopes and instruments, including the number of spectra and the time interval, are presented in Table 3. In addition to the BTA/MSS, echelle spectrographs were also employed. Most of the echelle spectra — twelve in total — were obtained with the BTA equipped with the NES spectrograph (Panchuk et al., 2017). The spectral resolution $R$ of the instruments, in Angstroms, was estimated by the *FWHM* of the O I 5577 Å night sky emission. All instruments operate in aperture mode, so the spectra are a combination of the star's spectrum and the spectrum of the night sky. High-resolution spectra were obtained primarily during the full Moon and sometimes with cirrus clouds. Under these conditions, the equivalent widths of the He II emission line tend to be underestimated due to the contribution of the night sky to the continuum. To correct these estimates, we used the equivalent widths of closely positioned check emission lines [Fe III] 4701.62 Å and Fe II 4731.44 Å, whose intensities are less variable.

In the high-resolution spectra, radial velocities were measured using the wind metal lines Fe I 6318.02 Å, Fe II 4555.89, 4629.34, 4731.44, 5316.70 (blend), 5534.86, 7865.56 Å, and also the forbidden nitrogen line [N II] 5754.59 Å. Among the iron lines, we selected the two to three strongest ones visible in a specific spectrum, and their radial velocities were averaged. The [N II] nitrogen line is unique in that its intensity did not change during the X-ray outburst of 1998, while the intensities of other lines in the spectrum increased tenfold. 250 days after the peak, the intensity of the [N II] line increased by a factor of 1.8 relative to the level observed during the outburst, and then dropped below the quiescent level (Barsukova et al., 2006b). This forbidden line can only form in the outer parts of the gas-and-dust envelope, where the gas density is very low. It was assumed that the intensity increase was due to the ejected matter from the explosion reaching the edges of the envelope. The ejection velocity was used to determine an upper limit for the CI Cam envelope radius of 200 AU. However, our observations show that the radial velocity of [N II], along with its profile shape and width in the quiescent state, do not differ significantly from those of the iron lines.

The X-ray spectrum of CI Cam shows a wide Fe XXIV–XXV $K_\alpha$ blend, in which Bartlett et al. (2013) identified neutral iron Fe I components. It is not clear to the authors how neutral and fully ionized iron atoms can exist simultaneously. These data may indicate that the compact object is immersed in circumstellar matter. We also observe the strong neutral iron line Fe I 6318.0171 Å in the optical spectrum, corresponding to the electronic transition between levels $3d^7(4F)4p \to 3d^6 4s^2$. This line is mistakenly identified as Fe II in some papers. It does not differ in radial velocity from other Fe II wind lines, having the same structure and width, and is clearly formed in the wind envelope of the B star.

The results of high-resolution spectroscopy, including radial velocities and full widths at half maximum for the iron and nitrogen lines, are shown in Figs. 3(f)–3(i) as a function of time.

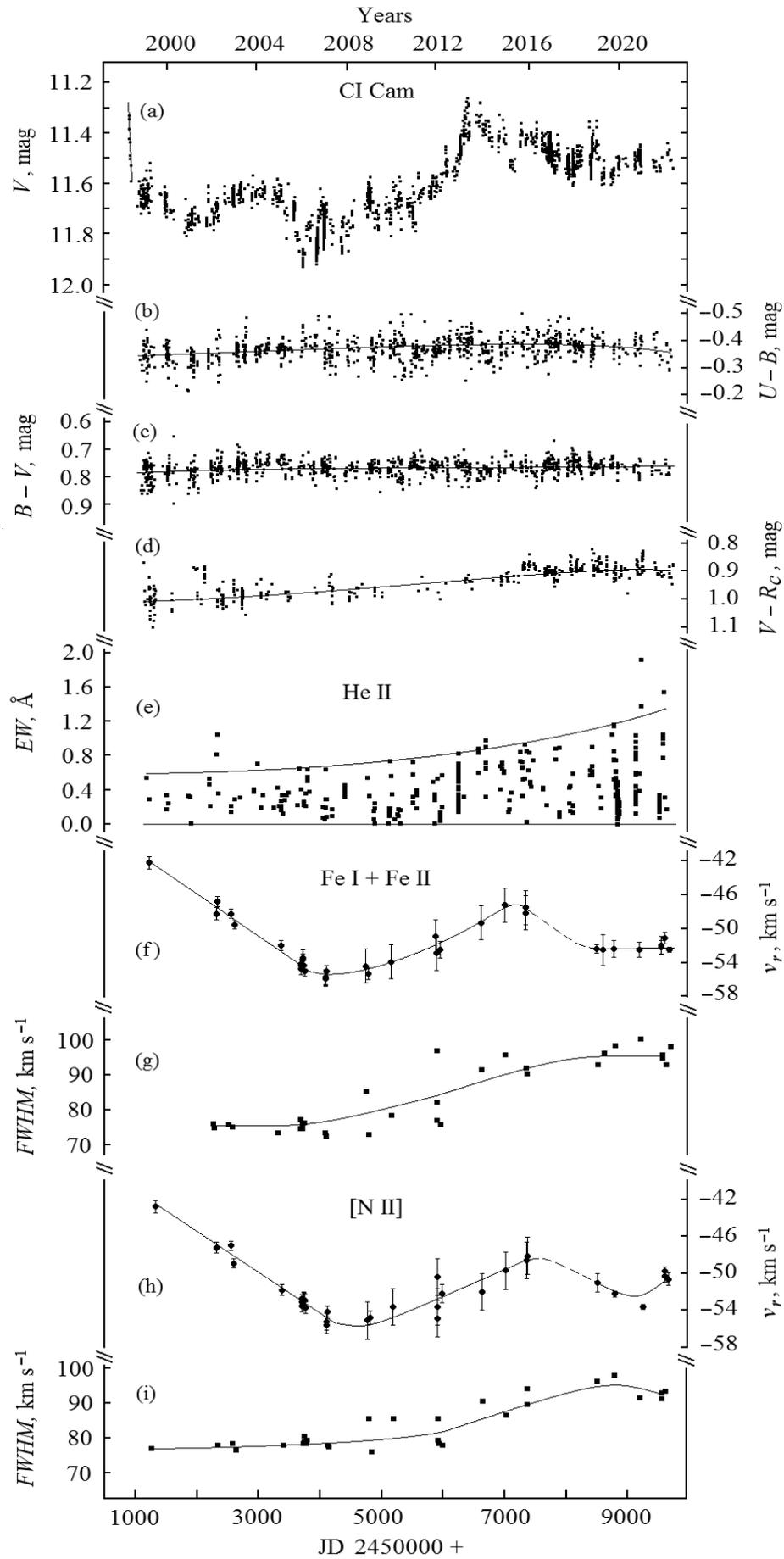

Fig. 3. Photometric and spectral variations of CI Cam in a quiescent state over time from 1998 to 2022. (a–d) V-band light curves and color indices U−B, B−V, and V−RC; (e) variations in the equivalent width of the He II 4686 Å emission; (f, g) variations in radial velocity and emission line widths at half maximum for iron lines; (h, i) variations in radial velocity and line widths for the forbidden line [NII] 5754.59 Å.

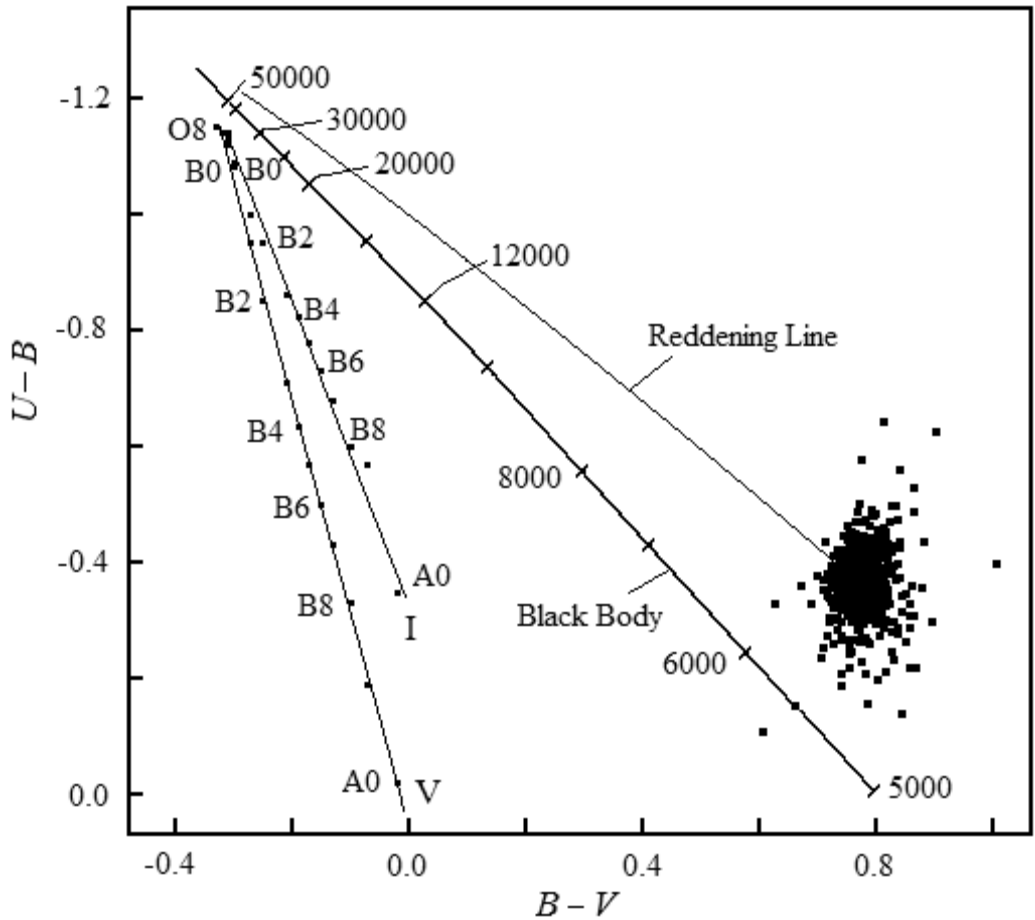

Fig. 4. *(U−B)−(B−V) color–color diagram. Color indices for normal main sequence B-stars (V) and supergiants (I) are plotted based on data from Straizys (1982), along with the computed blackbody dependence with indicated temperatures and the interstellar reddening line. The cloud of dots at the bottom right represents the color indices for the CI Cam continuum, with the contribution from emission lines subtracted, as determined from spectral data.*

**5. He II Emission Source and Its Orbit**

As evident from Fig. 3e, the equivalent width (*EW*) of the He II 4686 Å emission line fluctuates rapidly, ranging from zero to a maximum value, with this maximum value gradually increasing over the 24 years since the outburst.

Fig. 5 shows the results of spectral monitoring of the equivalent width of the He II 4686 Å line over five nights. Below the plots for this line, we present measurements of a fainter check line, [Fe III] 4701.62 Å, whose variation amplitude does not exceed the measurement errors. The average *EW* level for the He II line can vary by a factor of three from night to night. During a single night, the amplitude of variability can increase tenfold within half an hour.

The phase radial velocity curve for the He II 4686 Å emission line is plotted in Fig. 6a. The phase curve is constructed using the elements $T_{dn}$ = JD2455572.491+19.407E, where $T_{dn}$ is the moment of transit of the descending node of the orbit. The scatter in the radial velocity curve is significant, reaching up to 250 km s$^{-1}$ during the descending node phase. The full range of radial velocity variations exceeds 500 km s$^{-1}$ (corresponding to 8 Å in the spectrum). The emission line can be asymmetric, with multiple intensity peaks. The radial velocity was determined from the intensity-weighted average wavelength. The line can be faint, sometimes appearing against the continuum noise background. The large scatter in the radial velocity curve defines the accuracy of the period determination as ±0.004 days.

The warning from Bartlett et al. (2019) regarding the impossibility of using the He II 4686 Å emission 'to trace the dynamics' warrants special attention. The key question is whether the orbital elements of the He II emission source in CI Cam can be determined from the radial velocity of this emission? As an example of this line's unsuitability for orbital computations, Bartlett et al. (2019) cite the SS 433 microquasar and the work by Goranskij (2011). SS 433 is an eclipsing system with an orbital period of 13.082 days, where the eclipse is not total in the He II line profile. However, during the eclipse, the He II radial velocity curve passes through the zero point in a coordinate system of the mass center where the compact component is located at the exterior conjunction. This suggests that the line is formed precisely near the compact component or within its Roche lobe. Gies et al. (2002) found that the C II 7231, 7236 Å blend in SS 433 behaves exactly the same as the He II line. These lines of ionized elements have been used in both these studies and others ones to determine the orbit. Meanwhile, the radial velocities of the H and He I lines are maximal and directed away from the observer at the exterior conjunction of the compact object, passing through the zero point at orbital quadratures. In other words, their radial velocity curves are phase-shifted by $\pi/2$. These lines also form in the stellar wind but within a volume much larger than the size of the binary system (Gies et al., 2002). The compact object has a smaller mass, and the A4–A7 III star, which fills its Roche lobe ($q = 0.15$), blocks the hard emission from the compact component in a wide shadow cone and cannot ionize the hydrogen or helium atoms within it. Consequently, the profiles of these lines appear shifted towards the compact component due to the deficiency of radiating atoms in the shadowed region. As a result, the H and He I lines do not provide information about the orbital velocity but rather convey information about the component mass ratio $q$, on which the shadow's width depends. Conversely, He II is excited by the hard radiation from the collimated jets, whose structure does not obstruct the He II emission. The Doppler shift of this specific line allows for the estimation of the orbital velocity of the compact object.

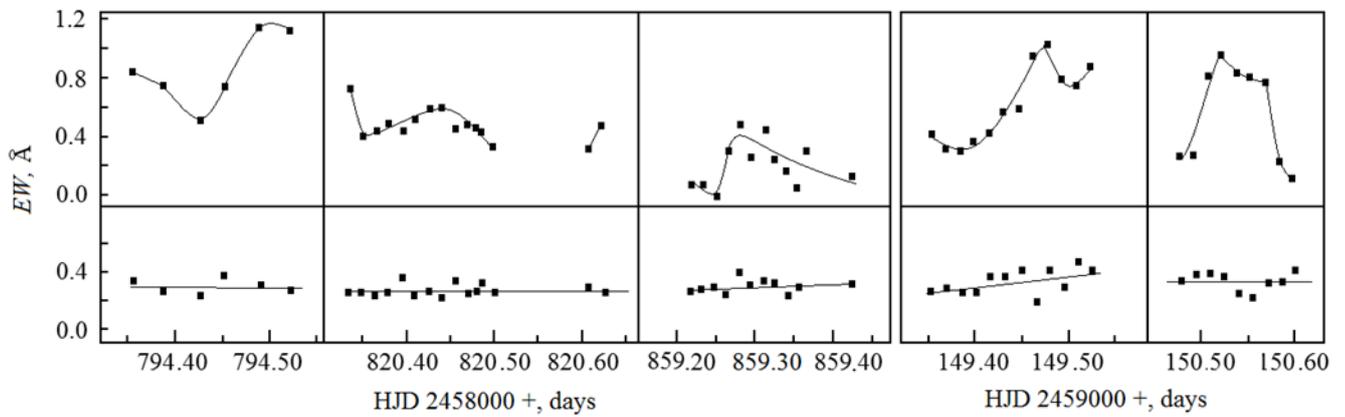

Fig. 5. *Rapid variations in the equivalent width of the He II 4686 Å line over the course of a night (upper parts of the plots) and equivalent width measurements of the nearby check line [Fe III] 4701.62 Å (lower parts) for five nights of observation.*

It is important to note that the Doppler shifts of the He II emission are reliable and often the only indicator of the radial velocity of a compact object in cataclysmic variables with hot white dwarfs, in classical nova remnants, and in symbiotic systems where this line is formed in the inner regions of accretion disks. In many cases, traces of a cool, low-mass component in the optical spectra of cataclysmic systems are simply absent.

In the case of CI Cam, the following should be noted:

(1) The B-type star dominating the optical region, along with its envelope formed by wind or a circumstellar disk, is not the source of either the emission or absorption of He II 4686 Å. The He II line profile differs from wind line profiles. In cases where the He II source emission disappears, no traces of the B-type star's atmospheric He II absorption are visible. The He II line is primarily observed in the spectra of hotter Of-type or O-type stars in absorption, although it can also be detected in the spectra of some Be-type stars, such as δ Sco (spectral type B0.3 IVe; Miroshnichenko et al., 2013). In this work, the orbit was determined from the He II line in absorption.

(2) He II 4686 Å is the only line in the optical spectrum of CI Cam that exhibits significant changes in its radial velocity over a wide range and with a clearly defined period. Therefore, it is associated with an object that is truly in motion.

(3) No influence of the motion of this object on the profiles and velocities of wind lines was found. Indeed, if the He II line was formed in the stellar wind from the disk around the compact component, we would observe the interaction between the winds of this disk and the B star. Barsukova and Goranskij (2009) reported the appearance of absorption components in the He I 4713 Å emission profile, which could be related to the propagation of pulsation waves in the B star's envelope. However, the connection between these features and the orbital motion has not yet been established. Additionally, the profile of this He I line differs from the rectangular profiles typically observed in the B star's wind lines.

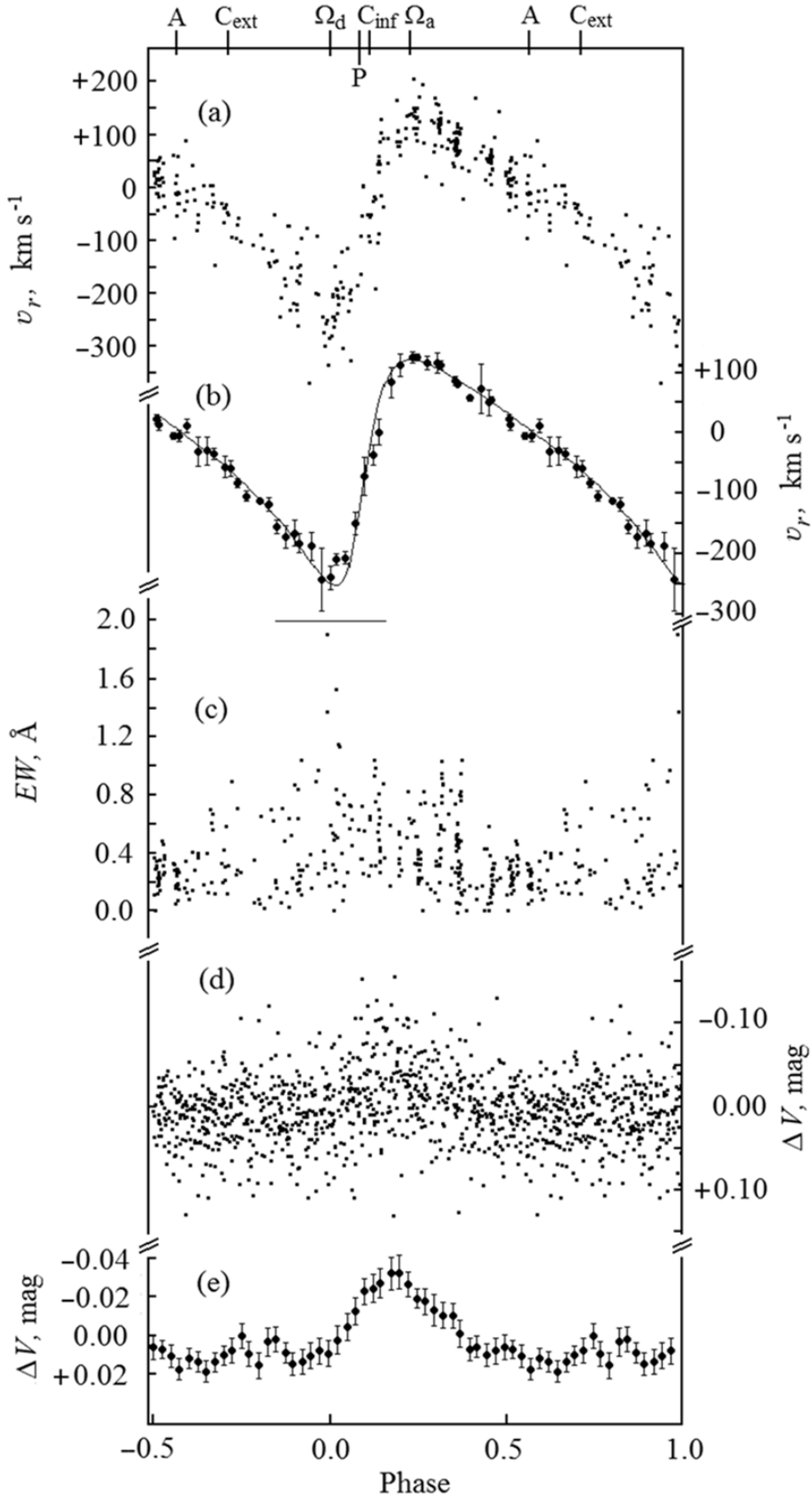

Fig. 6. *Spectral and photometric variations observed for CI Cam as a function of the 19.407-day orbital period phase: (a) radial velocity curve for the He II 4686 Å emission; (b) smoothed He II radial velocity curve. The solid line represents the model radial velocity curve constructed using graphically determined orbital elements; (c) He II line equivalent width versus orbital phase; (d) V-band phase light curve constructed from night-averaged observations. Slow brightness variations were*

*removed by Fourier decomposition into periodic components, followed by the subtraction of these components using the prewhitening procedure; (e) averaged light curve for the orbital period. Orbital phases are indicated at the top: A—apastron, $C_{ext}$—exterior conjunction of the He II emission source, $\Omega_d$—descending node (where the descending node marks the start of the orbital phase count), P—periastron, $C_{inf}$—inferior conjunction, $\Omega_a$—ascending node of the orbit.*

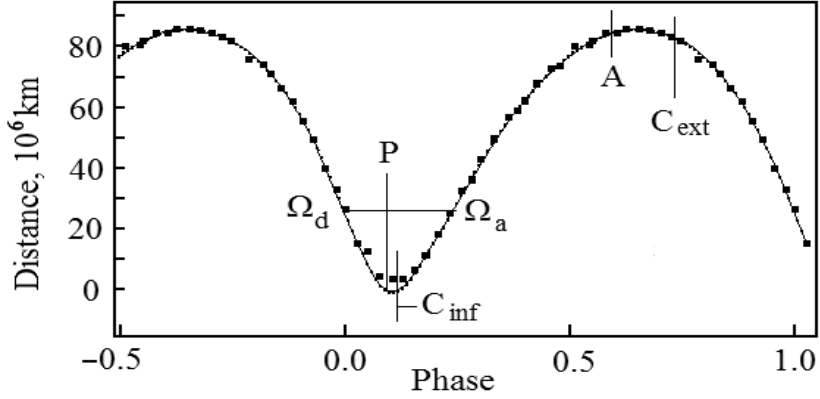

Fig. 7. *Integrated curve derived from the radial velocity curve, indicating the line-of-sight position of the He II source. The main computed orbital phases are marked, with their designations described in Fig. 6.*

All these arguments led us to conclude that the He II emission radial velocity can be used to determine the orbital elements of the secondary component and establish its nature. The orbital elements were determined from the average phase curve using the graphical method of Henroteau (1928). The radial velocity phase curve was constructed using the sliding average method (Fig. 6b). From this, the radial velocity semi-amplitude derived from the average curve is K = 190 km s$^{-1}$. Integrating the radial velocity curve over time (Fig. 7) yields the full amplitude of the He II emission source displacement along the line of sight to be 86 × 10$^6$ km, or 0.57 AU. The line-of-sight projected distance between the system's center of mass and the orbital periastron is 25 × 10$^6$ km, or 0.17 AU. This parameter defines an upper limit of 36 $R_\odot$ on the radius of the B star.

To estimate the accuracy of the orbital elements, we computed the orbit using individual measurements with the '*rvfit*' software from the IRAF package. A comparison of the elements derived from these two methods is provided in Table 4.

**Table 4.** A comparison of the orbital elements of CI Cam determined graphically and by a computational method

| Method | Graphical | Computational |
|---|---|---|
| Orbital period, days | 19.407 ± 0.004 | 19.4041 ± 0.0002 |
| Periastron passage epoch $T_P$, HJD | 2459028.266 | 2459028.52 ± 0.01 |
| Eccentricity $e$ | 0.49 | 0.44 ± 0.01 |
| Ascending node longitude $\omega$, degrees | 250 | 257.8 ± 0.9 |
| System velocity $\gamma$, km s$^{-1}$ | −33.5 | −33.0 ± 0.9 |
| Radial velocity semi-amplitude $K$, km s$^{-1}$ | 190 | 168.58 ± 0.02 |
| Mass function $f(M)$, $M_\odot$ | 8 | 7.0 ± 0.2 |
| Number of observations | 152 | 217 |

The dependence of the He II 4686 Å line equivalent width (*EW*) on the orbital phase is shown in Fig. 6c. The largest emission *EW* is observed near the descending node of the orbit; however, in this phase, the *EW* can also approach zero. Barsukova et al. (2021) explain the peak emission in this phase as resulting from the interaction between the emission source and the pulsation wave, where the total interaction surface is directed toward the observer, and the emission source has an extended structure. As the source enters the denser layers of the pulsating B star's envelope, it attains a relative velocity of more than 200 km/s with respect to the surrounding medium. The maximum *EW* values are reached over a wide range of orbital phases, from −0.1 to +0.4, during which the object is at periastron and inferior conjunction—i.e., within the densest regions of the B star's envelope. Values close to zero are rarer at inferior conjunction, between the orbital nodes, than at other phases. In phases 0.6–0.7 (near apastron), the scatter in *EW* decreases due to the absence of both extreme large and small values. Note that the decrease in *EW* occurs when the emission source is at its maximum distance from the B-type star, but not during the exterior conjunction phase of the source with the B star. Therefore, the weakening of the emission is not due to a total or partial eclipse of the source but is likely related to the emergence of the emission source into the less dense layers of the envelope near apastron.

Plots of *V*-band light curves folded with the orbital period, constructed separately from averaged brightness values for each night (a total of 786 nights), as well as the average light curve derived from these data using the sliding average method, are presented in Figs. 6d and 6e. The photometry results show that, on average, the brightness level of the star and the strength of the He II 4686 Å emission increase between the orbital nodes. At this point, the He II emission source passes near periastron, where the envelope density of the B star is at its maximum. The average amplitude of the brightness variations associated with orbital motion is 0.04 magnitudes.

## 6. B Star Pulsations

CI Cam is the only B[e]-type star that exhibits such clear and coherent pulsations (see Fig. 2). These pulsations were observed during seven monitoring campaigns for CI Cam, conducted between 2005 and 2020. Additionally, there is a series of continuous TESS satellite observations of this sky region, carried out from November 28 to December 23, 2019 (STScI MAST archive). The TESS light curve, converted to stellar magnitudes, is shown in Fig. 8. This light curve clearly shows both the orbital light variations and the pulsations. The orbital phases are indicated along the top edge of the figure. Thus, the pulsations of the B star and the orbit of the He II emission source are phenomena reliably confirmed by the most accurate observations conducted at an orbital space observatory operating in the optical and near-infrared (IR) range.

The connection between CI Cam's fast variability over the course of a night and its multi-periodic pulsations was discovered by Barsukova and Goranskij (2008). Oscillations in two modes, with periods of $P_1 = 0.4152$ days (amplitude = $0^m.020$) and $P_2 = 0.2665$ days (amplitude = $0^m.015$), dominated the spectrum (Barsukova and Goranskij, 2009). A detailed frequency analysis of the data from 2005–2006 was published by Goranskij and Barsukova (2009). The frequency spectrum was found to be complex, with seven periodic components identified. The third strongest component, with an amplitude of $0^m.012$, corresponds to the period $P_3 = 0.5223$ days. After the discovery in 2018 that CI Cam pulsates in a single mode with a period of 0.4062 days, close to $P_1$ (Goranskij and Barsukova, 2018), we conducted a frequency analysis of all earlier observations in monitoring mode, starting from 2012. We found that the star began pulsating exclusively in this mode from 2012 onward, coinciding with the increase in its brightness, which reached a maximum in 2013. Data from intensive monitoring in December 2012, obtained during the ascending branch of this brightness increase (marked by the letter "c" in Figs. 1 and 2), already exhibited pulsations in this single mode. This conclusion was further confirmed by the highly accurate TESS data (Fig. 2g). The amplitude spectrum of the TESS data series shows no periodic components with periods shorter than one day, aside from this single oscillation. A significant scatter in the light curve is attributed to irregular brightness changes on a timescale of more than one day. The orbital component has been removed from the TESS data shown in Fig. 2g.

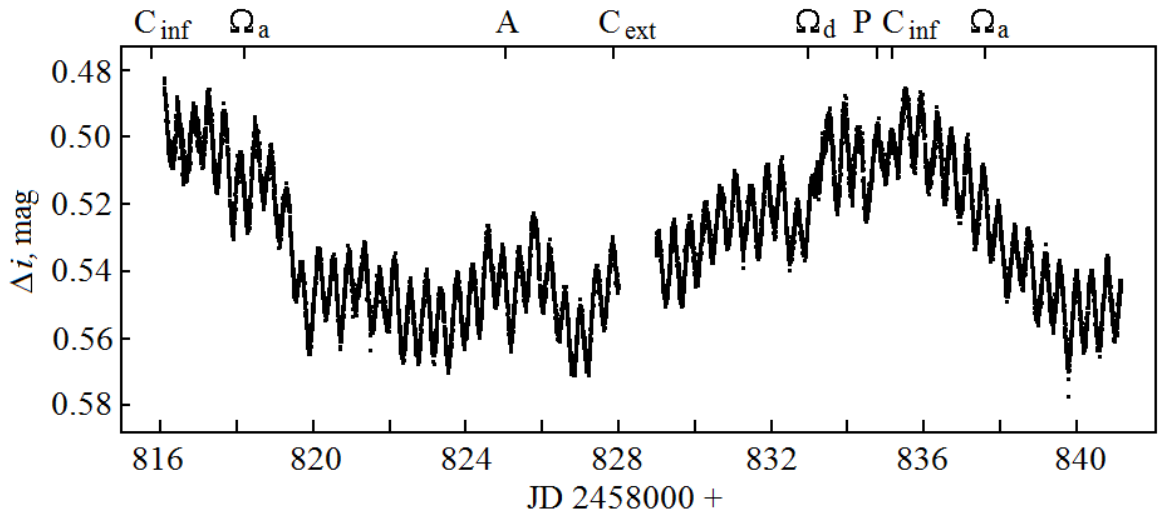

Fig. 8. *CI Cam light curve from TESS satellite observations in November–December 2019. Pulsations with a period of 0.4056 days and orbital light variations with a period of 19.407 days are observed simultaneously. The computed orbital phases are marked at the top, with the designations consistent with those in Fig. 6.*

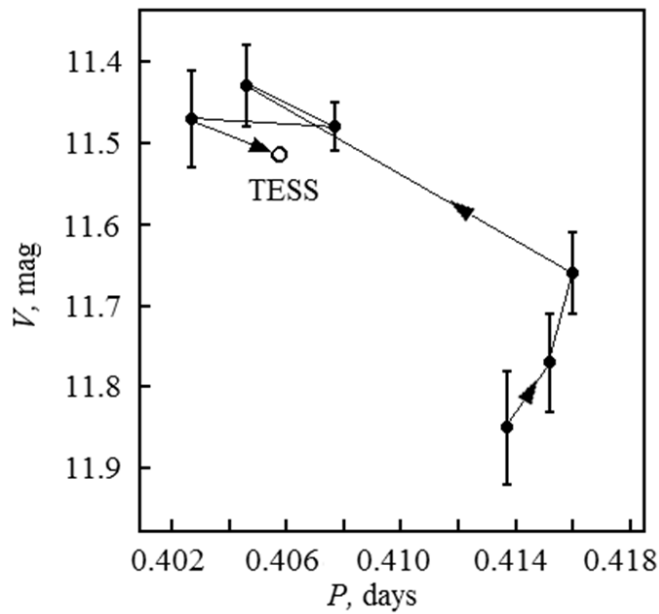

Fig. 9. *"First overtone period—average magnitude" dependence for the pulsations of CI Cam. TESS satellite data are indicated by a circle.*

Fig. 9 shows the dependence of the period $P_1 \approx 0.41$ days on the average $V$-band magnitude of CI Cam. The star follows a track on this diagram, indicating that the pulsation period gradually increases with the star's brightness and then decreases in a jump-like manner. The period decreases by 0.014 days, which is 3.4% of the total value, corresponding to a brightness increase of $0^m.4$ in the $V$ filter. The accuracy of the period determination varied from 0.0015 to 0.0026 days (0.38-0.65%), depending on the duration of the monitoring campaigns, which ranged from 13 to 22 days.

Pulsations are widespread among B- and Be-type stars. Two types of pulsating stars are known within spectral type B: β Cep and SPB (Aerts, 2021). These stars are pulsating B-type stars near the main sequence (MS). β Cephei stars, named after the prototype β Cephei, are early B-type stars that pulsate in p-modes (pressure modes) with periods ranging from 3 to 6 hours. SPB (Slowly Pulsating B-type) stars, which are B-type stars of intermediate subtypes, are photometric variables that pulsate in g-modes (gravity modes) with periods of several days (De Cat, 2002). Hybrid B-type stars have also been discovered, demonstrating simultaneously two types of pulsations. The full range of masses for pulsating main sequence (MS) B-type stars spans from 3 to 23 solar masses ($M_\odot$). The distribution of B-type stars by pulsation frequency exhibits two peaks corresponding to these star types, with a wide gap between them (De Cat, 2002). Moreover, the pulsation frequencies of CI Cam fall into that gap.

High precision data analysis of O- and B-type stars obtained by the TESS satellite (Burssens et al., 2020) has shown that even O5 Ifc to B9 Iab supergiants pulsate, exhibiting SLF[3] and SPB types of pulsations. However, the pulsations of CI Cam are coherent on timescales of months to years and do not belong to these types. Owocki and Cranmer (2002) numerically studied radial and non-radial pulsations originating in the winds of hot OB-type stars, as well as disturbances in their envelopes caused by stellar pulsations. Spectral and photometric studies of such connections and interactions are critical for understanding CI Cam.

However, it should be noted that the pulsation periods of CI Cam are typical of RR Lyrae-type stars. RR Lyrae stars have masses less than that of the Sun and pulsate in radial modes. The period $P_3$ is characteristic of RRab-type stars, which primarily pulsate in the fundamental mode. The period ratio $P_1 / P_3 = 0.795$ for CI Cam is close to the 3:4 period ratio observed in RRd-type stars with double-mode pulsations in the fundamental (F) and first overtone (1H) modes. The period ratio $P_2 / P_3 = 0.510$ in CI Cam is close to 2:1, a ratio observed in RR Lyrae-type stars exhibiting the Blazhko effect. Borkowski (1980) explained the Blazhko effect in AR Her by suggesting that, in addition to the fundamental mode (F) wave, another wave with a period about half that of the fundamental mode is resonantly excited (with a ratio close to 2:1). This additional wave may correspond to either the second or third overtone (2H or 3H).

The period ratio for oscillations with the highest amplitude in CI Cam, $P_2 / P_1 = 0.642$, is close to 2:3. To date, based on data from OGLE (Optical Gravitational Lensing Experiment), as well as the space missions COROT and Kepler, intensive studies of RR Lyrae-type stars in globular clusters and Cepheids in the Magellanic Clouds have led to the discovery of a large number of stars whose frequency spectra exhibit a signal with a period $P_x$ or $P_{0.61}$. This results in a period ratio with the first overtone $P_x/P_{1H}$ ranging from 0.60 to 0.65 (see Jurcsik et al., 2015; Moskalik et al., 2015; Netzel et al., 2015; Soszyński et al., 2010, and references therein). A wave with such a period ratio typically has a low amplitude, only 5% of the amplitude of the first overtone. Moreover, such a wave is observed only in stars where the first overtone is dominant. Some researchers consider the wave with a period $P_x$ as a non-radial mode, even emphasizing this suggestion in a paper title ("Double-mode Radial—Non-radial RR Lyrae Stars...", Netzel et al., 2015). There have also been discoveries of rare three-mode RR Lyrae-type stars, such as EPIC 201585823 — a star that pulsates in the fundamental mode, the first overtone, and additionally in a 'nonradial mode', with a period ratio of 0.616285 to the period of the first overtone (Kurtz et al., 2016). Several such stars have already been discovered in globular clusters and the Galactic bulge. Netzel and Smolec (2022) published the results of asteroseismological computations for RR Lyrae-type stars and Cepheids with triple modes, using a code for a pulsating envelope. Waves with period ratios of 0.61 to the first overtone are explained by nonradial mode harmonics of orders 8 or 9 for RR Lyrae-type stars and 7, 8, or 9 for classical Cepheids.

Taking all these studies into account, we can reliably identify the wave with a period of $P_1 = 0.4152$ days in CI Cam as a first overtone pulsation (1H) and the wave with a period of $P_2 = 0.2665$ days as a pulsation in the mysterious 'nonradial mode,' which, with a small amplitude, is characteristic of many pulsating stars. However, significant objections remain regarding the non-radial nature of this wave.

Goranskii (1989) confirmed Borkowski's suggestion that the Blazhko effect in RR Lyrae-type stars is due to a 2:1 resonance of two radial modes, using the example of the RR Lyrae-type star AH Cam. Goranskij et al. (2010) showed that double-mode pulsations (F and 1H) in RR Lyrae-type stars are also a resonance, but with a period ratio close to 3:4. In V79, located in the globular cluster M3, two states of double-mode pulsations were observed: the dominant 1H mode (double-mode pulsations, resonance 3:4) and the dominant F mode (Blazhko effect, resonance 3:4). The amplitude modulation for such a Blazhko effect with a 3:4 resonance had a very short period $1^d.389$. Later a classical Blazhko effect with a 2:1 resonance was also observed in the same star with an amplitude modulation period of $65^d.4$.

The following are signs of radial mode resonance:
- Waves in the light curve with period ratios close to whole-number ratios: 1:2, 2:3, 3:4.
- The secondary wave amplitude depends on the phase of the dominant wave.
- Non-additivity of these waves in the overall light curve: the model light curve, consisting of the sum of the average light curves of the isolated waves, has a significantly lower amplitude than the real, observed light curve.

---

[3] SLF—a stochastic low-frequency process—random, unpredictable light variations which can be described by one mathematical parameter.

The 'beating' of two waves in resonance appears as a long-term energy transfer from one wave to the other when the wave phases are close and coincide, followed by an energy drain when they are in antiphase. Such a resonance is known in engineering as parametric. In a system where the oscillation period depends on a specific parameter (e.g., the pendulum oscillation period depends on its length), resonance is caused by periodic action directed at that parameter (the pendulum length). (The mechanism of oscillation build-up in a child's swing). For a star, this parameter is its radius $R$. For energy transfer, the wave interaction period ratio needs to be close to a ratio of whole numbers. As shown by Borkowski (1980), the period of the resonance mode $P_x$ is determined by the period of the fundamental mode $P_F$ and the light curve modulation period $P_B$ (Blazhko effect period) as $P_x = 1/(2/P_F + 1/P_B) \approx P_F/2$.

The mechanism of radial resonance between the fundamental mode (F) and the first overtone (1H) in stars is well understood. Motion in the F mode involves all the layers of the stellar envelope moving in phase. In contrast, motion in the 1H mode involves the upper and lower layers moving in antiphase, with a stationary layer — the node between them. In double-mode pulsations (F and 1H), the node of the first overtone moves along the radius with the period of the fundamental mode. The zones of helium and hydrogen ionization, located beneath the surface of the star, act as the driving mechanism of stellar pulsations. When a pulsating star expands, helium recombination occurs, and the accompanying energy release provides a boost to the expanding envelope. If the recombination zone coincides with the position of the first overtone node, the boost energy is transferred into the motion of the fundamental mode. As the recombination zone moves away from the node, the energy sustains the motion of the first overtone. The process becomes more complex if the envelope pulsates in two overtones. However, parametric resonance occurs only in systems with two or more eigenfrequencies (pulsation modes) if there is no external periodic influence on them. Nonetheless, resonances between the eigenfrequencies have not yet been reproduced using the pulsating envelope model.

When researching the nature of CI Cam, a question arises: which stars, besides RR Lyrae-type stars, can pulsate with periods close to those discovered by the photometric method? According to the Eddington formula for the pulsation constant, derived for self-gravitating spheres of gas, $Q = P\sqrt{\rho}$, where $P$ is the period in days and $\rho$ is the average density of the star in solar units, the radial pulsation period depends only on the average density of the star. Christy (1966) used nonlinear computations for RR Lyrae-type pulsating stars to calculate the pulsation constants. He obtained $Q_F = 0.0365$ for the fundamental mode and $Q_F = 1.34 Q_{1H}$ for the relation between the pulsation constants of the fundamental mode and the first overtone. This gives $Q_{1H} = 0.0272$. It follows from the Eddington formula that:

$$\rho = \left(\frac{Q}{P}\right)^2$$

Substituting the periods of the first overtone of CI Cam and the pulsation constant of the first overtone, taking into account the average density of the Sun, we derive the average density of CI Cam as 0.0060 g cm$^{-3}$ and 0.0063 g cm$^{-3}$ before and after the brightness increase in 2013, respectively. That is, the observed decrease in the period of the first overtone with increasing brightness is related to the decreasing radius of the star. Similar computations for the fundamental mode period of 0.5223 days yield an average density of CI Cam of 0.0069 g cm$^{-3}$ (the difference is about 10%).

Let us compare the information obtained from the pulsation data with the constraints on mass and radius derived from spectral data and orbital parameters (see the 'mass–radius' diagram in Fig. 10). The constraints are as follows: the lower mass limit for the B star is 8 $M_\odot$, and its upper radius limit is 36 $R_\odot$.

Let us modify the Eddington formula:

$$\rho = \frac{M}{\frac{4}{3}\pi R^3} = \left(\frac{Q}{P}\right)^2$$

Therefore

$$M = \frac{4}{3}\pi R^3 \left(\frac{Q}{P}\right)^2$$

The position of radially pulsating stars with a first overtone period of 0.4152 days, in logarithmic scales of log $M$ versus log $R$, can be represented by a straight line log $M = 3 \log R - 1.745$. Fig. 10 shows this straight line along with the positions of normal O9–B5-type stars belonging to luminosity classes V (main sequence), III (giants), and Iab (supergiants). Masses, radii, and other parameters for normal stars are taken from the book by Straizys (1982). Stars of spectral types B0–B2 III are positioned along this straight line. These stars have masses ranging from 12 to 22 $M_\odot$ and radii between 8.3 and 11 $R_\odot$. Based on these data, the absolute magnitudes $M_v$ of such stars range from $-3^m.7$ to $-4^m.9$, color indices $(B-V)_0$ range from $-0^m.24$ to $-0^m.30$, and $(U-B)_0$ range from $-0^m.88$ to $-1^m.09$. B0–B3 Iab supergiants still fall within the elliptical orbit but cannot exhibit radial pulsations with these periods. Figure 10 shows the upper limit on radial pulsations, which passes through O9 V, O9 III, and O9 Iab stars. These are hot stars where the helium ionization region is located on the surface and, therefore, cannot be responsible for the pulsation-sustaining mechanism in the envelopes.

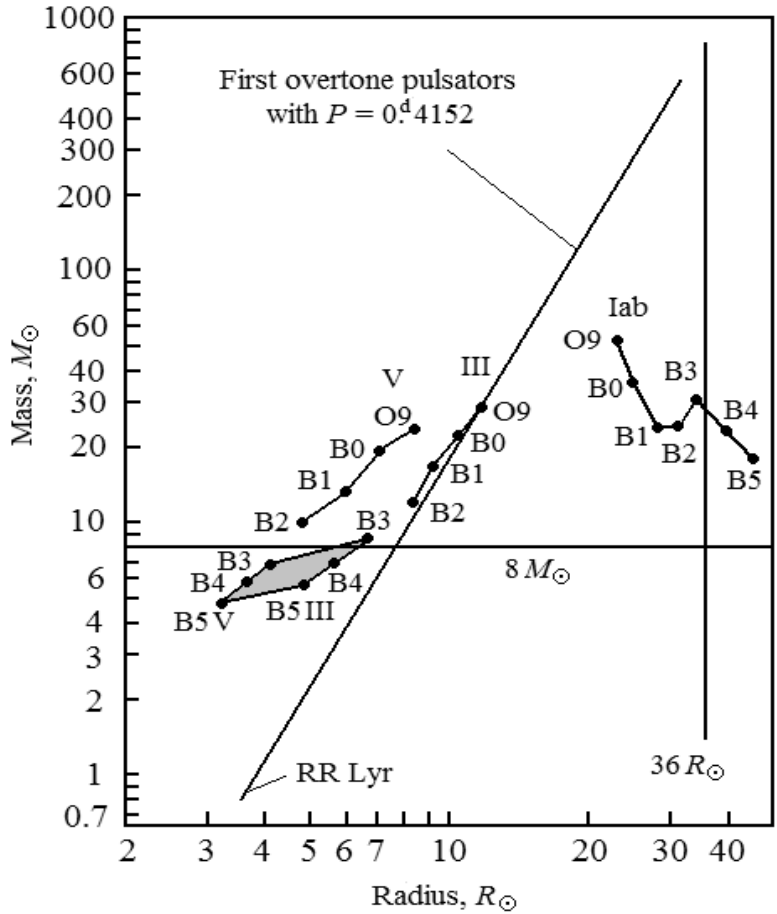

Fig. 10. *Logarithmic scale 'mass–radius' diagram for normal stars from the main sequence (V) to supergiants (Iab), with limits for the main CI Cam system component: 8 $M_\odot$ in mass and 36 $R_\odot$ in radius (indicated by the horizontal and vertical lines, respectively). The inclined line represents the position of pulsating stars in the first overtone with the period shown in the figure. The grey rhombus highlights the region corresponding to the B4 III–V classification of CI Cam, based on the hydrogen line absorption width.*

At the same time, we can conclude that the B component in the CI Cam system is positioned deep inside the orbit of the He II emission source, with the periastron of the orbit located at least at a height equal to two of its radii. Moreover, pulsation waves propagate in the envelope at such a height, allowing the He II emission source to interact with them.

Even if we assume the maximum absolute magnitude for a pulsating B0 III star, such as CI Cam, to be $-4^m.9$ mag, with a maximum registered $R$-band brightness of $7^m.1$ and an outburst amplitude of $3^m.5$, the resulting absolute magnitude at the outburst peak would be $-8^m.4$. This value is insufficient for identifying CI Cam as a supernova impostor.

## 7. Acceleration in Wind Lines and Its Nature

Slow secular radial velocity changes in the iron emission lines, and even in the forbidden nitrogen line, present another mystery of CI Cam. Such a phenomenon has not yet been observed in other B[e]-type stars. As evident from Figure 3 (curves f and h), the linear trend of decreasing velocity reversed in 2007. Goranskij et al. (2017) explained this phenomenon as the result of a third component passing close to the periastron of the elliptical orbit, leading to an increase in the brightness of the B-type star, changes in its pulsation characteristics, and alterations in the orbit of the He II emission source. The radial velocity curve of the B star could be approximated by an orbital solution with a lower period limit of 220 years. During the discussion of this report at the 2016 conference on B[e]-type stars in Prague (Czech Republic), Steven N. Shore proposed an alternative interpretation of this phenomenon — a precession of the B star's rotation axis and its polar outflows. New high-resolution spectral observations trace the increasing velocity up to 2015. However, after a three-year gap in observations, the velocity decreased again. This same trend is observed in the iron lines Fe I and Fe II, as well as in the forbidden nitrogen line [N II].

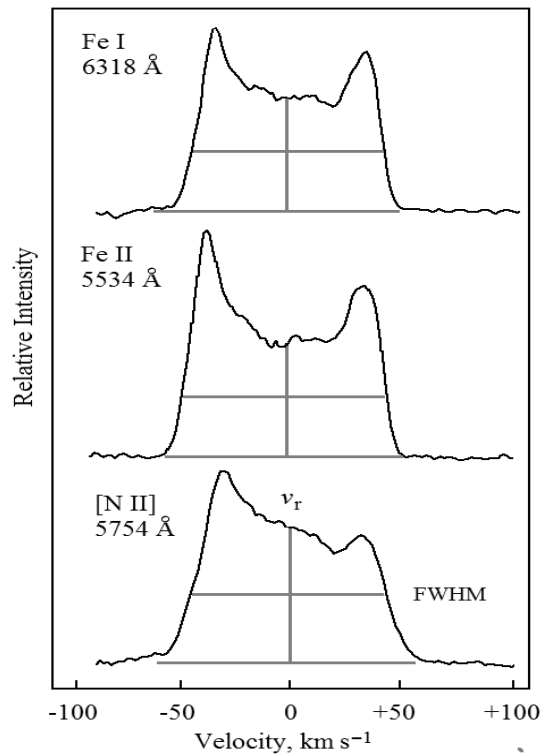

Fig. 11. *Profiles of the iron and forbidden nitrogen emission lines in the spectrum of CI Cam, obtained with the CFHT equipped with the ESPaDOnS echelle spectrograph on January 16, 2018. The method for determining the radial velocity $v_r$ and line widths (FWHM) is illustrated.*

Rectangular wind lines with flat tops and steep profile boundaries are often observed in classical novae and Wolf–Rayet stars. These profiles typically form in optically thin lines within spherical envelopes, where atoms are accelerated by radiation pressure from the central star (Beals, 1931). While other geometric configurations, as well as variations in density and velocity distributions, could also produce such profiles (Williams and Mason, 2010), these factors are generally less influential. Additionally, the profiles of permitted lines may exhibit optical depth effects. In giant stars, the disk of the star can obscure parts of the expanding envelope that are moving away from the observer at maximum velocity. These effects are evident in the wind line profiles of CI Cam (Fig. 11). The tops of the wind line profiles are not flat but instead exhibit two peaks: one high peak at the blue edge and another lower peak at the red edge. Additionally, a gradual decrease in intensity from the blue to the red edge is observed in the profile. This phenomenon is related to an increase in absorption and light scattering along the line of sight to the observer. Additionally, the CI Cam envelope is denser compared to those in classical novae. Although the [N II] emission region is optically thin, emission from the opposite side of the envelope must pass through the optically thick inner regions, where it is partially scattered. This explains the difference in the structure of this emission line compared to iron lines.

The effect of optical depth and the overlapping of receding stellar wind flows by the star's disk is evident in the discrepancy between the systemic velocity of CI Cam ($\gamma = -33$ km s$^{-1}$), determined from the radial velocity curve of the He II 4686 line, and the average velocity of the B star (vr = $-52$ km s$^{-1}$), as determined from the radial velocity of the Fe II and [N II] wind lines. This discrepancy amounts to $-19$ km s$^{-1}$. Evidently, this effect also influences the full width at half maximum (*FWHM*) of the iron and nitrogen wind lines. Due to the optical depth effect, the velocity variations of the B star, as measured by the emission line profiles, are influenced by changes in the wind envelope transparency and by contributions from wind flows moving away from the observer. At the maximum B star velocity, measured immediately after the outburst (vr = $-42$ km s$^{-1}$; Figs. 3f and 3h), even the photospheric spectrum of the star was observed (Barsukova et al., 2006; Hynes et al., 2002), which later disappeared as the envelope transparency decreased. Stellar wind density variations in the CI Cam envelope occur on a time scale of several years. This constitutes the third hypothesis explaining the acceleration observed in the wind lines.

The light scattering effect in the dense envelope medium can become apparent if a He II emission source moves through it, causing the total light of the system to increase by $0^m.04$ in the *V* filter (Figs. 6d and 6e) and by $0^m.05$ in the *i* filter (Fig. 8) when passing through periastron. In this case, the contribution of this source to the star's continuum spectrum can be significant. At the same time, the source is not eclipsed by the B star.

To test these hypotheses, we measured the full width at half maximum (*FWHM*) of the iron line and the forbidden nitrogen line. Evidently, precession of the outflows at the two poles of the star would lead to an expansion or contraction of the line profile. For the *FWHM* measurements, we used the intensity midway between the two emission line peaks as an upper limit (even if it is lower than the red peak), and the width was determined at the midpoint of this intensity. Samples of the iron and nitrogen line profiles, as well as the methods for measuring *FWHM* and radial velocity, are shown in Fig. 11. The results of the *FWHM* measurements are presented in Fig. 3g for the Fe lines and in Fig. 3k for the [N II] line. During the observation period, the width of the "rectangular" lines increased from 74 to 94 km s$^{-1}$. Taking into account the optical depth effect and the obscuring of the stellar wind by the disk of the B star, this corresponds to a wind velocity change from 56 to 66 km s$^{-1}$. The intensification of the wind is clearly related to the increase in the star's brightness in 2013.

Unfortunately, our high-resolution spectral observations do not yet allow us to interpret the acceleration phenomenon in the wind lines conclusively. There is a lack of high-resolution spectra for the period between 2012 and 2018. The periodicity of this phenomenon has not yet been confirmed. The first observations with the BTA/PFES pertain to the period of activity just after the 1998 outburst and, therefore, may distort the possible periodic pattern. Although the orbit of the third component, with a period of over 220 years, is not confirmed, we do not rule out a shorter 14-year period. The relationship between the radial velocities of the B star and the stellar wind density can be examined through model computations of the wind envelope.

## 8. Results Discussion

Thus, optical observations of pulsations allow us to refine the parameter limits for the primary CI Cam system component, such as its spectrum within the range B0–B2 III, its mass, and its radius, independently of other observed data. The intrinsic $(B-V)_0$ color indices for B0–B2 III stars on the color–color diagram (Fig. 4) fall within a narrow interval of $-0^m.24$ to $-0^m.30$. Since the average CI Cam continuum color index, excluding contributions from emission lines, is $+0^m.78$, the color excess can be determined without an accurate $U-B$ color index for the B star continuum. The color excess is thus $1^m.05 \pm 0^m.03$, while the interstellar reddening of CI Cam is $A_V = 3^m.23 \pm 0^m.09$. Adopting a brightness for CI Cam of $V=11^m.55$ in its quiescent state, and considering the range of absolute magnitudes for B0–B2 III stars to be between $-3^m.7$ and $-4^m.9$, we derive a distance of 2.5 to 4.5 kpc. These distance limits are consistent with the value from the latest Gaia data release (EDR3), which is 4.1 (−0.2/+0.3) kpc.

In addition to the spectral type limits B0–B2 III for the B star and the mass range derived from pulsation data, we also have the mass function from the spectral data of the He II emission source. We adopt the orbital inclination angle reported by Thureau et al. (2009), who discovered a dust disk around the B star and considered it coplanar with the component's orbit. We then substitute the previously determined parameters into the mass function equation for the component:

$$f(M_1) = \frac{M_2^3 \sin^3 i}{(M_1 + M_2)^2}$$

where $f(M_1)$ is the mass function of 7-8 $M_\odot$, $M_2$ is the mass of the B star (12–22 $M_\odot$), $i$ is the orbital inclination angle (67°), and $M_1$ is the unknown mass to be solved for. For a B0 III star, we derive the component mass to be within 10–12 $M_\odot$, while for a B2 III star, the mass ranges from 0.98 to 1.77 $M_\odot$.

The low-mass component in the CI Cam system is likely a white dwarf, and the 1998 outburst can be attributed to a thermonuclear explosion of hydrogen accumulated on its surface due to acn from the B star's envelope. This hypothesis constrains the spectral type and mass of the primary component to B2 III and 12 $M_\odot$, respectively. There are at least ten arguments supporting this hypothesis.

1. The X-ray spectrum of CI Cam during the outburst, as observed by BeppoSAX and ASCA, is optically thin and thermal up to 10 keV. It also contains many high-ionization emissions (K-shell) from elements such as O, Ne, Si, S, and Fe (Frontera et al., 1998; Orr et al., 1998; Ueda et al., 1998). Transients involving black holes or neutron stars do not produce such spectra. (Ishida et al., 2004).
2. The X-ray spectrum is characterized by a two-temperature thermal distribution, and the duRn of the outburst is approximately one week (Orlandini et al., 2000). This duration is consistent with computations for a thermonuclear explosion on the surface of a white dwarf with a mass of 1 $M_\odot$ (Iben, 1982).
3. The soft X-ray component resembles that of SSS-type sources, such as CAL 87 in the LMC, which are characterized by thermonuclear burning on the surface (Ishida et al., 2004).
4. There are no signs of fast variability or a hard tail component in the X-ray spectrum, which are characteristic of black holes, nor are there pulsations typically seen in neutron stars (Orlandini et al., 2000).
5. In the quiescent state, the X-ray spectrum exhibits emission from an optically thin plasma with $kT_2$ =5.5 keV. This emission is similar to that observed in white dwarfs accreting in dwarf nova systems (DN); however, in dwarf novae, the luminosity is an order of magnitude lower than that of CI Cam, given a distance to CI Cam of 5–17 kpc (Ishida et al., 2004).
6. During the CI Cam outburst, an expanding envelope was observed in the radio range, which is attributed to the ejection of helium- and hydrogen-rich layers resulting from a thermonuclear explosion. This is similar to the envelopes expelled during classical nova explosions (Orlandini et al., 2000). Such envelopes have not been observed in X-ray transients involving neutron stars or black holes.
7. During the CI Cam outburst, the ejected envelope was visible in the line profiles as a pedestal. Its expansion velocity, along with its deceleration parameters, matches the velocity observed in the radio range. No broad emission line components were detected that would suggest the presence of a jet, as seen in the symbiotic star Z And.
8. In the quiescent state, there are no distortions in the wind line profiles, even during periastron passage. This observation suggests that the component has a low mass.
9. The large semi-amplitude of radial velocities (190 km s$^{-1}$) observed for the He II emission source, in contrast to the absence of such velocities in the stellar wind lines of the B-type star, provides additional support for the low mass of the component.
10. The absolute magnitude at the peak of the outburst, after subtracting the contribution from the B[e] star, is on average $M_R = -8^m.7$ (with error margins due to distance inaccuracies ranging from $-7^m.9$ to $-9^m.2$). In comparison, the brightest classical novae in M31 have absolute magnitudes $M_V$ ranging from $-8^m.5$ to $-9^m.9$ during an outburst (Della Valle, 2002).

The low-mass component in the CI Cam system, the He II emission source, is at a later stage of evolution, as it is a compact object. This indicates that the initial mass transfer has already occurred. At the start of its evolution, this component was a more massive star, which passed through the red supergiant phase. Its massive, hydrogen-rich envelope was transferred to a less massive component, which is now the massive B[e] star. Some of the matter may not have reached the smaller component (a non-conservative scenario) but instead formed an envelope around the system, where circumstellar dust was created. According to models by van Rensbergen et al. (2008), the amount of such matter could be several solar masses. From the results of our pulsation analysis, the mass of the primary component, the B-type star, now amounts to about 12 $M_\odot$. The B[e]-type star is surrounded by a relatively dense, expanding envelope and possibly a circumstellar gas disk. The B star is moving away from the main sequence (MS) towards the red giant region. At the same time, the star's volume has exceeded that of its Roche lobe, and a reverse mass exchange has begun. This is evidenced by the intense stellar wind and the accumulation of hydrogen-rich material on the surface of the low-mass component, as well as the thermonuclear hydrogen explosion that led to the star's outburst in 1998, which was observed across all regions of the electromagnetic spectrum. There is no clear understanding of how the secondary component remains on an elliptical orbit after the initial mass exchange. This may possibly be related to the presence of a third component in the system.

The assumption about the compactness of the low-mass component of the system also requires verification. Occasionally, we detect a sharp increase in He II emission during the orbital descending node passage. At the same time, the component enters the dense layers of the B star's envelope with a velocity exceeding 200 km s$^{-1}$, and the portion of its surface that collides with the circumstellar medium faces the observer. A similar effect could occur if the component is surrounded by a thick accretion disk, resulting in a collision between the surrounding gas and the disk's edge surface. However, we do not observe stable emission from such a disk in the spectrum.

The behavior of the optical light curve, which shows an increase in system brightness during inferior conjunction, suggests that the component makes a significant contribution to the total system spectrum. Part of its emission is likely scattered and absorbed in the envelope or circumstellar disk of the B star. This contribution can be estimated by constructing a model of a circumstellar nebula formed by the stellar wind and the light absorption within it. However, a lower limit for the luminosity of the low-mass component can be estimated as 100–300 $L_\odot$ based on the 0.04-mag reduction in the total $V$-band brightness of the system, which includes a B0-B2 III-type star. Such a high luminosity contradicts the white dwarf hypothesis. As a result of the first mass exchange, the component may be a remnant of a helium core from a star that lost its hydrogen envelope during the initial mass exchange and now radiates due to central helium burning. It may turn out to be a hot Of-type subdwarf, whose spectrum cannot be distinguished from the B star's emission in the total spectrum. The 1998 CI Cam outburst can then be explained by a hydrogen flare-up in a layer at the base of the subdwarf's hydrogen envelope, which had accumulated as a result of accretion from the B star's envelope.

## 9. CONCLUSIONS

We have obtained light curves and color indices for CI Cam in the $UBVR_C$ system over a 24-year period following the 1998 outburst. We note an increase in the system's brightness by an average of 0$^m$.4 in 2013.

The spectrum of the B[e]-type star exhibits rapid variations in the equivalent width of the He II 4686 Å emission line on a timescale of tens of minutes. The orbital parameters of this emission source have been refined, with the orbital eccentricity determined to be between 0.43 and 0.49. The dependence of the emission intensity on the orbital phase, as well as its variation over the 24 years since the 1998 outburst, was investigated.

We investigated the pulsations of the primary component of CI Cam (a B-type star), which we had discovered earlier. Between 2005 and 2009, these pulsations were multiperiodic and could be described by a resonance involving three waves: the fundamental mode, the first overtone, and a wave that forms a period ratio of 0.64 with the first overtone. The latter wave is frequently observed in RR Lyrae-type stars and Cepheids with the first dominant overtone. All three waves are identified as radial pulsations. Following the brightness increase in 2013, the star began to pulsate solely in the first overtone. A "period–brightness" dependence was observed, indicating that as the brightness level increased, the radius of the star decreased.

Based on the average density determined from pulsation data, the primary component of the CI Cam system can be classified as a B0–B2 III-type giant with a mass ranging from 12 to 22 $M_\odot$. The suggestion that the He II emission source is a white dwarf with a mass less than 1 $M_\odot$, combined with the known mass function, constrains the classification of the primary component to a B2 III-type giant with a mass of 12 $M_\odot$ and a radius of 8.3 $R_\odot$. The classification of CI Cam as an sgB[e] star is entirely ruled out based on the observed pulsation periods. It is likely that CI Cam is a system that has undergone its first mass transfer and can be identified as a FS CMa-type B[e] star. Therefore, unresolved questions remain for future investigations.

## 10. DIGITAL ARCHIVE OF OBSERVATIONS

Observations of CI Cam are available in open access, stored as a ZIP archive and in ASCII file format at: https://relay.sao.ru/jet/~bars/CICam-ARCHIVE/
(https://relay.sao.ru/hq/bars/CICam-ARCHIVE/ARCHIVES.zip).
The archive contains tables with photometric data, as well as spectral data including radial velocities and equivalent line widths discussed in this paper.

**ACKNOWLEDGMENTS**

The authors are grateful to N.V. Borisov, A.F. Valeev, G.G. Valyavin, A.S. Vinokurov, V.V. Vlasyuk, V.G. Klochkova, D.N. Monin, V.E. Panchuk, S.N. Fabrika, M.V. Yushkin and S.Yu. Shugarov, who took part in the observations, showed an interest in this investigation, and provided observational material for analysis. This work makes use of the Gaia DR1–DR3 databases, TESS (Mikulski Archive for Space Telescopes at STScI), and the NIST Atomic Spectra Database. Observations with the SAO RAS telescopes are supported by the Ministry of Science and Higher Education of the Russian Federation. The renovation of telescope equipment is currently provided within the national project "Science and Universities". The work was performed as part of the SAO RAS government contract approved by the Ministry of Science and Higher Education of the Russian Federation. We thank the Russian Telescope Time Allocation Committee for providing the BTA observing time, and also the administration of SAO RAS for the Zeiss-1000 telescope observing time. V.P. Goranskij and E.A. Barsukova are grateful to the administrations of MSU, SAI and the Crimean Astronomical Station of SAI MSU for the allocated Crimean station telescope time and longstanding support of the work on this subject. This paper is partly based on observations obtained at the Canada-France-Hawaii Telescope (CFHT), which is operated by the National Research Council of Canada, the Institute National des Sciences de l'Univers of the Centre National de la Recherche Scientifique de France, and the University of Hawaii. The observations at CFHT were performed with care and respect from the summit of Maunakea, which is a significant cultural and historic site.

**FUNDING**
P.L.N., L.H.I. and K.A.S. would like to thank the Bulgarian Research Foundation for partial financial support of this work with a bilateral grant KP-06-RUSSIA-9/2019. The study was financially supported by the Russian Foundation for Basic Research and the National Science Foundation of Bulgaria as a part of the scientific project No. 19-52-18007 and Grant KP-06-Russia-9/2019. S.Z. and A.M. acknowledge PAPIIT grants IN102120 and IN119323. This research has been funded in part by the Science Committee of the Ministry of Education and Science of the Republic of Kazakhstan
(Grant No. AP08856419).

**CONFLICT OF INTEREST**
The authors declare no conflict of interest.

**REFERENCES**
1. C. Aerts, Rev. Modern Physics **93** (1), article id. 015001 (2021).
2. V. L. Afanasiev and A. V. Moiseev, Astronomy Letters **31** (3), 194 (2005).
3. D. A. Allen and J. P. Swings, Astron. and Astrophys. **47**, 293 (1976).
4. E. A. Barsukova, N. V. Borisov, A. N. Burenkov, et al., Astronomy Reports **50** (8), 664 (2006a).
5. E. A. Barsukova, N. V. Borisov, A. N. Burenkov, et al., ASP Conf. Ser. **355**, 305 (2006b).
6. E. A. Barsukova, N. V. Borisov, V. P. Goranskii, et al., Astronomy Reports **46** (4), 275 (2002).
7. E. A. Barsukova, A. N. Burenkov, and V. P. Goranskij, Astronomer's Telegram, No. 14362 (2021).
8. E. A. Barsukova, S. N. Fabrika, S. A. Pustil'nik, and A. V. Ugryumov, Bull. Spec. Astrophys. Obs. **45**, 147 (1998).
9. E. A. Barsukova and V. P. Goranskij, Astronomer's Telegram, No. 1381 (2008).

10. E. A. Barsukova and V. P. Goranskij, Comm. Asteroseismology **159**, 71 (2009).
11. E. A. Barsukova, V. G. Klochkova, V. E. Panchuk, et al., Astronomer's Telegram, No. 1036 (2007).
12. E. S. Bartlett, J. S. Clark, M. J. Coe, et al., Monthly Notices Royal Astron. Soc. **429** (2), 1213 (2013).
13. E. S. Bartlett, J. S. Clark, and I. Negueruela, Astron. and Astrophys. **622**, id. A93 (2019).
14. C. S. Beals, Monthly Notices Royal Astron. Soc. **91**, 966 (1931).
15. T. Belloni, S. Dieters, M. E. van den Ancker, et al., Astrophys. J. **527** (1), 345 (1999).
16. Y. K. Bergner, A. S. Miroshnichenko, R. V. Yudin, et al., Astron. and Astrophys. Suppl. **112**, 221 (1995).
17. K. J. Borkowski, Acta Astronomica **30**, 393 (1980).
18. S. Burssens, S. Simón-Díaz, D. M. Bowman, et al., Astron. and Astrophys. **639**, id. A81 (2020).
19. S. Carpano, F. Haberl, C. Maitra, and G. Vasilopoulos, Monthly Notices Royal Astron. Soc. **476** (1), L45 (2018).
20. R. F. Christy, Astrophys. J. **144**, 108 (1966).
21. J. S. Clark, A. S. Miroshnichenko, V. M. Larionov, et al., Astron. and Astrophys. **356**, 50 (2000).
22. J. S. Clark, I. A. Steele, R. P. Fender, and M. J. Coe, Astron. and Astrophys. **348**, 888 (1999).
23. P. De Cat, ASP Conf. Ser. **259**, 196 (2002).
24. T. J. Deeming, Astrophys. and Space Sci. **36** (1), 137 (1975).
25. M. Della Valle, AIP Conf. Proc. **637**, 443 (2002).
26. R. A. Downes, Publ. Astron. Soc. Pacific **96**, 807 (1984).
27. F. Frontera, M. Orlandini, L. Amati, et al., Astron. And Astrophys. **339**, L69 (1998).
28. D. R. Gies, M. V. McSwain, R. L. Riddle, et al., Astrophys. J. **566** (2), 1069 (2002).
29. V. P. Goranskii, Sov. Astron. **33**, 45 (1989).
30. V. Goranskij, Peremennye Zvezdy **31** (5), 5 (2011).
31. V. Goranskij and E. A. Barsukova, Peremennye Zvezdy **7** (15), 15 (2007).
32. V. Goranskij, C. M. Clement, and M. Thompson, in *Proc. Intern. Conf. on Variable Stars, the Galactic halo and Galaxy Formation, Zvenigorod, Russia, 2009*, Ed. by C. Sterken, N. Samus, and L. Szabados (Astronomical Institute of Moscow University, Moscow, 2010), p. 115.
33. V. P. Goranskij and E. A. Barsukova, Astrophysical Bulletin **64** (1), 50 (2009).
34. V. P. Goranskij and E. A. Barsukova, Astronomer's Telegram, No. 12097 (2018).
35. V. P. Goranskij, E. A. Barsukova, K. S. Bjorkman, et al., ASP Conf. Ser. **508**, 307 (2017).
36. M. Heida, R. M. Lau, B. Davies, et al., Astrophys. J. **883** (2), article id. L34 (2019).
37. A. Henden and U. Munari, Astron. and Astrophys. **458** (1), 339 (2006).
38. F. C. Henroteau, Handbuch der Astrophysik **6**, 299 (1928).
39. R. M. Hjellming, A. J. Mioduszewski, E. L. Robinson, et al., IAU Circ., No. 6862, 1 (1998).
40. R. I. Hynes, J. S. Clark, E. A. Barsukova, et al., Astron. and Astrophys. **392**, 991 (2002).
41. I. Iben Jr., Astrophys. J. **259**, 244 (1982).
42. M. Ishida, K. Morio, and Y. Ueda, Astrophys. J. **601** (2), 1088 (2004).
43. G. H. Jacoby, D. A. Hunter, and C. A. Christian, Astrophys. J. Suppl. **56**, 257 (1984).
44. J. Jurcsik, P. Smitola, G. Hajdu, et al., Astrophys. J. Suppl. **219** (2), 25 (2015).
45. V. G. Klochkova, V. E. Panchuk, and M. V. Yushkin, Astrophysical Bulletin **77** (1), 84 (2022).
46. A. Kostenkov, A. Vinokurov, Y. Solovyeva, et al., Astrophysical Bulletin **75** (2), 182 (2020).
47. D. W. Kurtz, D. M. Bowman, S. J. Ebo, et al., Monthly Notices Royal Astron. Soc. **455** (2), 1237 (2016).
48. H. J. G. L. M. Lamers, F.-J. Zickgraf, D. de Winter, et al., Astron. and Astrophys. **340**, 117 (1998).
49. J. F. Le Borgne, G. Bruzual, R. Pelló, et al., Astron. and Astrophys. **402**, 433 (2003).
50. P. W. Merrill and C. G. Burwell, Astrophys. J. **78**, 87 (1933).
51. A. J. Mioduszewski and M. P. Rupen, Astrophys. J. **615** (1), 432 (2004).
52. A. S. Miroshnichenko, Astronomical and Astrophysical Transactions **6** (4), 251 (1995).
53. A. S. Miroshnichenko, Astrophys. J. **667** (1), 497 (2007).
54. A. S. Miroshnichenko, V. G. Klochkova, K. S. Bjorkman, and V. E. Panchuk, Astron. And Astrophys. **390**, 627 (2002).
55. A. S. Miroshnichenko, A. V. Pasechnik, N. Manset, et al., Astrophys. J. **766** (2), article id. 119 (2013).
56. P. Moskalik, R. Smolec, K. Kolenberg, et al., Monthly Notices Royal Astron. Soc. **447** (3), 2348

(2015).
57. H. Netzel and R. Smolec, Monthly Notices Royal Astron. Soc. **515** (3), 3439 (2022).
58. H. Netzel, R. Smolec, and P. Moskalik, Monthly Notices Royal Astron. Soc. **453** (2), 2022 (2015).
59. M. Orlandini, A. N. Parmar, F. Frontera, et al., Astron. and Astrophys. **356**, 163 (2000).
60. A. Orr, A. N. Parmar, M. Orlandini, et al., Astron. and Astrophys. **340**, L19 (1998).
61. S. P. Owocki and S. R. Cranmer, ASP Conf. Ser. **259**, 512 (2002).
62. V. E. Panchuk, V. G. Klochkova, and M. V. Yushkin, Astronomy Reports **61** (9), 820 (2017).
63. E. L. Robinson, I. I. Ivans, and W. F. Welsh, Astrophys. J. **565** (2), 1169 (2002).
64. D. Smith, R. Remillard, J. Swank, et al., IAU Circ., No. 6855, 1 (1998).
65. I. Soszyński, R. Poleski, A. Udalski, et al., Acta Astronomica **60** (1), 17 (2010).
66. V. Straizys, *Metal-deficient stars* (Institut Fiziki Akad. Nauk Litovskoj SSR, Mokslas, Vil'nyus) (1982).
67. N. D. Thureau, J. D. Monnier, W. A. Traub, et al., Monthly Notices Royal Astron. Soc. **398** (3), 1309 (2009).
68. Y. Ueda, M. Ishida, H. Inoue, et al., Astrophys. J. **508** (2), L167 (1998).
69. W. van Rensbergen, J. P. De Greve, C. De Loore, and N. Mennekens, Astron. and Astrophys. **487** (3), 1129 (2008).
70. V. A. Villar, E. Berger, R. Chornock, et al., Astrophys. J. **830** (1), 11 (2016).
71. V. Šimon, C. Bartolini, A. Guarnieri, et al., New Astronomy **12** (7), 578 (2007a).
72. V. Šimon, C. Bartolini, A. Guarnieri, et al., Open European Journal on Variable Stars **0075**, 24 (2007b).
73. R. Williams and E. Mason, Astrophys. and Space Sci. **327** (2), 207 (2010).